\documentclass[aps,twocolumn,floatfix,showpacs,amssymb]{revtex4}
\usepackage{graphicx,bm}
\usepackage{amsmath}
\begin{document}
\title{Isotropic contact forces in arbitrary representation:\\
Heterogeneous few-body problems and low dimensions}
\author{Ludovic Pricoupenko}
\affiliation
{Laboratoire de Physique Th\'{e}orique de la Mati\`{e}re Condens\'{e}e, 
Universit\'{e} Pierre et Marie Curie and CNRS, 4 place Jussieu, 75252 Paris, France.}
\date{\today}
\begin{abstract}
The Bethe-Peierls asymptotic approach which models pairwise short-range forces by contact conditions is introduced in arbitrary representation for spatial dimensions less than or equal to 3. The formalism is applied in various situations and emphasis is put on the momentum representation. In the presence of a transverse harmonic confinement, dimensional reduction toward two-dimensional (2D) or one-dimensional (1D) physics is derived within this formalism. The energy theorem relating the mean energy of an interacting system to the asymptotic behavior of the one-particle density matrix illustrates the method in its second quantized form. Integral equations that encapsulate the Bethe-Peierls contact condition for few-body systems are derived. In three dimensions, for three-body systems supporting Efimov states, a nodal condition is introduced in order to obtain universal results from the Skorniakov Ter-Martirosian equation and the Thomas collapse is avoided. Four-body bound state eigenequations are derived and the 2D '3+1' bosonic ground state is computed as a function of the mass ratio.
\end{abstract}
%
%03.65.Nk Scattering theory
%03.65.Ge Solutions of wave equations: bound states
%21.45.+v Few-body systems
%34.50.-s Scattering of atoms and molecules
%34.10.+x General theories and models of atomic and molecular collisions and interactions (including statistical theories, transition state, stochastic and trajectory models, etc.) 
%05.30.Jp Boson systems (for static and dynamic properties of Bose-Einstein condensates, see 03.75.Hh and 03.75.Kk)
%05.30.Fk Fermion systems and electron gas (see also 71.10.-w Theories and models of many-electron systems; see also 67.10.Db Fermion degeneracy in quantum fluids)
%04.20.Cv Fundamental problems and general formalism
%32.80.Pj Optical cooling of atoms; trapping
\maketitle

\section{Introduction}

The non-perturbative zero-range limit of a pairwise interaction was introduced in the context of nuclear physics for the description of the deuteron \cite{Bet35,Bre47,Bla52}. In this modeling, the finite range pairwise interacting potential is replaced by a zero-range potential and  a peculiar asymptotic behavior is imposed on the wave function as the two interacting particles approach each other. More than 60 years after these pioneering works, the Bethe-Peierls approach appears to be especially relevant for modeling pairwise interaction in few- and many-body systems of ultra-cold atoms where details of the interaction are useless for understanding a large class of low-energy processes \cite{Ols98,Pet00,Pet01,Pet03,Pet04a,Cas04,Pri04a,Idz06,Pet05,Pea05,Wer06a,Wer06b,Kar07a,Kar07b}. This remarkable simplicity follows essentially from two features: first, the temperature in these systems is sufficiently low that ${s}$-wave scattering is the dominant process in two-body collisions \cite{pwave}; second, the three-dimensional (3D) $s$ wave scattering length (denoted ${a_3}$) in the two-body scattering can be tuned by use of a magnetic Feshbach resonance (FR) toward an arbitrary large value with respect to the range of the interatomic forces (denoted by ${b_3}$) and which is of the order of the van der Waals radius \cite{Fes62,Chi10}: 
\begin{equation}
b_3 \sim \left(\frac{\mu C_6}{\hbar^2}\right)^{1/4} ,
\end{equation}
where ${C_6}$ is the van der Waals coefficient of the interatomic potential and $\mu$ is the reduced mass of the two interacting particles. In the vicinity of a broad FR the two-body scattering cross section ${4\pi a_3^2}$ is very large with respect to the non-resonant cross section ${(\sim 4\pi b_3^2)}$ and this justifies the use of the 3D Bethe-Peierls approach only parameterized by ${a_3}$.

Ultra-cold atoms in reduced geometries are the subject of intensive experimental and theoretical studies \cite{QGLD}. Decisive progress have been obtained both in quasi-1D systems (for example with the observation of the Tonks Girardeau gas \cite{Kin04a,Par04}) and in quasi-2D systems. A general review of low-dimensional systems can be found for example in Refs.~\cite{Blo08,Had09}. Systems in reduced dimensions are achieved by using very anisotropic trapping potentials leading to a $D$-dimensional behavior where ${D=1}$ (quasi-1D systems) or ${D=2}$ (quasi-2D systems). In an ideal $D$-dimensional atomic waveguide, noninteracting atoms move freely along $D$ direction(s) while they remain frozen in the lowest state of the transverse zero-point motion characterized by a length denoted as ${a_{\rm trans}}$. The associated energy ${E_{\rm trans} = \hbar^2/(2\mu a_{\rm trans}^2)}$ defines the limit of energy beyond which higher transverse states are populated and a 3D-like behavior is progressively recovered for increasing energies. In the actual experimental state-of-the-art, atomic waveguides have a transverse atomic length which is large as compared to the 3D potential radius ${(b_3)}$. Therefore for low energy processes \emph{i.e.}, for collisional energies $E$ such that ${|E| \ll \hbar^2/(\mu b_3^2)}$, the short-range pairwise interacting potential can be described through the 3D Bethe-Peierls approach. For collisional energies much smaller than ${E_{\rm trans}}$ the transverse excited states of the trap are populated only \emph{via} virtual processes and for large interatomic separations $r$ ${(r \gg a_{\rm trans})}$ only the ground transverse state is occupied. In this regime of collisional energies and in the free $D$-dimensional subspace, scattering process can be deduced from a $D$-dimensional effective pairwise interaction characterized by a finite range denoted ${b_D}$, which is of the order of the transverse length ${a_{\rm trans}}$. Hence for low energy $D$-dimensional processes where ${|E| \ll \hbar^2/(\mu b_D^2) }$ the effective low-dimensional interaction can be analogously to the 3D case, replaced by a zero-range force. Therefore in each dimension ${D}$ less than or equal to 3, a ${D}$-dimensional Bethe-Peierls approach can be relevant and useful as a tool for exploring the properties of few- and many-body shallow states. The dimensional reduction of the effective interaction from 3D to 1D was first achieved in Ref.~\cite{Ols98} and from 3D to 2D in Ref.~\cite{Pet00}.

One of the purpose of this paper is to show that the zero-range potential approach can be handled in a very simple and unified way in any dimension ${D \le 3}$. The key tool used along these lines is introduced in Sec. \ref{sec:tools} of the paper. It consists of the general expression of the Bethe-Peierls asymptotic condition which can be used in any representation for all dimensions ${D}$. Substitution of the ${\delta}$-source terms of the zero-range approach by a family of well-behaved functions of vanishing but finite support allows one to avoid technical problems concerning double-limit calculations (\emph{i.e.}, evaluation of infinite series in the zero-range limit). While used usually in the configuration space, the zero-range approach can thus be implemented directly in the momentum representation for translation invariant systems. The link between this formulation of the Bethe-Peierls approach and the $\Lambda$ potential introduced in Ref.~\cite{Ols02} is analyzed. The method is illustrated in Sec. \ref{sec:Dreduction} with the example of the dimensional reduction issue from 3D toward 1D and 2D. The Bethe-Peierls contact condition is directly expressed in the functional basis which diagonalizes the free Hamiltonian and known results are obtained in a simple manner. For $N$-body systems, integral equations which encapsulate the Bethe-Peierls contact condition are obtained in Sec. \ref{sec:Nbody}. It is also shown how to express this zero-range force approach in a second quantized representation. As an example, energy theorems \cite{Ols03,Tan08a,Tan08b,Com09,Wer10} are derived by using this last formulation. Section-\ref{sec:fewbody} of the paper is devoted to the three- and four-body shallow bound state issue. For three identical bosons in 3D, the original Skorniakov Ter-Martirosian (STM) equation \cite{Sko57} appears as a straightforward application of the results of the preceding section. A nodal condition is imposed on the three-body wave function in order to avoid the Thomas collapse \cite{Tho35} while keeping the simplicity of the zero-range approach \cite{Pri10b}. Known results of universal theory \cite{Bra06} are recovered from the regularized STM equation. For two-mass component fermionic systems in 3D, critical mass ratios for the Efimov threshold are given in each partial wave by using the STM equation. Finally, the formalism is applied to derive four-body equations in the zero-range limit.  In 2D, the eigenenergy equation for two-mass component tetramers is studied numerically. The energy of the ground $s$ wave of '3+1' bosonic tetramers in the zero-range limit is computed as a function of the mass ratio.

In this paper the norm of any vector ${\mathbf v}$ is denoted $v$. For a uni-dimensional system (${D=1}$), ${\mathbf v}$ is algebraic and ${v=|{\mathbf v}|}$ is the absolute value of ${\mathbf v}$.

\section{Isotropic Contact Forces}

\label{sec:tools}

\subsection{Two-body transition matrix and scattering amplitude}

This section reviews basic notions of two-body scattering in $D$-dimensional spaces which are useful in the formulation of the Bethe-Peierls approach.

Two colliding particles are described in their center-of-mass frame by the relative particle of reduced mass ${\mu}$ and relative momentum ${\mathbf k}$. They are supposed to interact through a short range potential ${V}$, and the Hamiltonian for the relative particle is ${H = H_0 + V}$, where ${H_0}$ is the free Hamiltonian. In what follows, as a consequence of the small collisional energy, scattering is supposed to only occur in the $s$ wave channel of the relative particle. For an incoming atomic wave of momentum ${{\mathbf k}_0}$ and of collisional energy ${E= \hbar^2 k_0^2/2\mu}$ (measured from the continuum  threshold), the scattering state ${|\Psi_{\mathbf k_0}\rangle}$ of the relative particle verifies the Lippmann-Schwinger equation:
\begin{equation}
|\Psi_{\mathbf k_0}\rangle = |\mathbf k_0 \rangle + \frac{1}{E+i0^{+} -H_0}  V |\Psi_{\mathbf k_0}\rangle.
\label{eq:Lippmann-Schwinger}
\end{equation}
We now turn to the ${\mathbf k}$ representation with the convention 
$\langle{\mathbf r}|{\mathbf k}\rangle = \exp(i{\mathbf k}\cdot{\mathbf r})$:
\begin{equation}
\langle {\mathbf r} | \Psi  \rangle 
= \int \frac{d^D{k}}{(2\pi)^D} \langle {\mathbf k} | \Psi \rangle \exp(i{\mathbf k}\cdot {\mathbf r}).
\end{equation}
In the ${\mathbf k}$ representation, the scattering states in Eq.~\eqref{eq:Lippmann-Schwinger} can be written as a function of the half on-shell transition matrix (or $t$-matrix) defined by:
\begin{equation}
\langle \mathbf k |T(E+i0^{+}) | \mathbf k_0 \rangle = \langle \mathbf k | V |\Psi_{\mathbf k_0}\rangle ,
\end{equation}
so that:
\begin{equation}
\langle {\bf k} | \Psi_{{\bf k_0}} \rangle = (2\pi)^D \delta({\bf k} - {\bf k}_0) +  
\frac{\langle \mathbf k |T(E+i0^{+})|\mathbf k_0 \rangle}{E + i 0^{+} - \frac{\hbar^2k^2}{2\mu}} .
\label{eq:scatt_state}
\end{equation}
The pairwise interaction is supposed to be short range with a typical radius ${b_D}$. For a small energy ${|E| \ll \hbar^2/(\mu b_D^2)}$ and momentum ${|{\bf k}| b_D \ll 1}$, the half on-shell $t$-matrix only depends on the collisional energy $E$ and coincides with the low energy on-shell $t$-matrix denoted as ${T_D}$:
\begin{equation}
\langle \mathbf k |T(E+i0^{+})|\mathbf k_0 \rangle \simeq  T_D(E+i0^{+}) .
\label{eq:tmatrix}
\end{equation}
At the lowest order in energy (implying that ${|k_0| b_D \ll 1}$), the expression of $T_D$ can be parameterized by the $D$-dimensional scattering length $a_D$. It can be written as \cite{Sto88,Gir04}:
\begin{equation}
T_D(E+i0^{+}) = \frac{\Omega_D \hbar^2}{2\mu} \times 
\left\{
\begin{array}{ll}
\frac{\displaystyle a_3}{\displaystyle(1 +  i a_3 k_0)}  & \ (D=3)\\
\\
\frac{\displaystyle -1}{\displaystyle\ln(-i a_2 k_0 e^\gamma/2)} & \ (D=2) \\
\\
\frac{\displaystyle -ik_0}{(\displaystyle 1 + i a_1 k_0)} & \ (D=1) ,
\end{array}
\right.
\label{eq:TD}
\end{equation}
where ${\Omega_D}$ is the full $D$-dimensional space angle:
\begin{equation}
{\Omega_3=4\pi},\  {\Omega_2=2\pi}, \ \mbox{and} \ {\Omega_1=2}.
\end{equation}
In Eq.~\eqref{eq:TD} for the two-dimensional case, $\gamma$ is the Euler's constant, and the 2D scattering length $a_2$ is always positive. The transition matrices in Eq.~\eqref{eq:TD} are the basic objects for describing low energy scattering processes in $D$-dimensional few- and many-body systems. Scattering properties can be equivalently described with the $D$-dimensional scattering amplitudes defined by:
\begin{eqnarray}
&&f_3(k_0)=-a_3/(1+ik_0a_3) ,  \label{eq:f3} \\
&&f_2(k_0)=1/\ln(-ik_0a_2 e^\gamma/2) ,\label{eq:f2}\\
&&f_1(k_0)=-1/(1+ik_0a_1). \label{eq:f1}
\end{eqnarray}
The notion of scattering amplitude is often used in the configuration space where the scattering states for large relative coordinates ${(r \gg b_D)}$ are deduced from  Eq.~\eqref{eq:TD}:
\begin{equation}
\langle {\bf r} | \Psi_{{\bf k_0}} \rangle =e^{i \mathbf k_0 \cdot \mathbf r}  +  f_D(k_0) \times
\left\{
\begin{array}{ll} 
\frac{\displaystyle e^{ik_0 r}}{\displaystyle r} & (D=3) \\
\\
\frac{i \pi}{2} H_0^{(1)}(k_0r) &  (D=2)\\
\\
e^{ik_0r} &  (D=1) .
\end{array}
\right.
\label{eq:diff_r}
\end{equation}
In Eq.~(\ref{eq:diff_r}), for $D=2$ the outgoing scattering wave function $H_0^{(1)}$ is the Hankel's function of order zero \cite{Convention_f2D}; for $D=1$, $f_1$ is also called the even scattering amplitude \cite{Ols98}.

\subsection{Bethe-Peierls asymptotic approach in arbitrary representation}

Surprisingly the zero-range approximation has been essentially used in the configuration space while other representations can yield substantial simplifications. For example, as a consequence of translation invariance, the momentum representation is very well suited for solving the few-body problem in homogeneous space. In this section, it is shown how the zero-range approximation can be formulated in arbitrary representation.

While for finite range forces Eq.~\eqref{eq:tmatrix} is only valid for a small relative momentum $k$ (\emph{i.e.}, for ${k b_D \ll 1}$), in the zero-range potential approximation this equality is extended for arbitrary large values of $k$. That way, the short range pairwise potential $V$ is replaced by a formal zero-range potential such that the scattering states in this approximation coincide with Eq.~\eqref{eq:scatt_state} in the small relative momentum limit (${k b_D \ll 1}$) or with Eq.~\eqref{eq:diff_r} for large interparticle distances (${r\gg b_D}$). The simplest way to implement the zero-range approximation is to follow the Bethe-Peierls method \cite{Bet35}, where for any state $|\Psi\rangle$ described by the Hamiltonian ${H_0+V}$, the pairwise short range potential $V$ is replaced by a delta-source 
term ${|\delta^D \rangle}$ with an amplitude $S_\Psi$:
\begin{equation}
V |\Psi \rangle \longrightarrow  S_\Psi | \delta^D \rangle ,
\label{eq:substitution}
\end{equation}
and $S_\Psi$ is deduced from an asymptotic condition on the state $|\Psi\rangle$. In Eq.~\eqref{eq:substitution}, ${| \delta^D \rangle}$ is such that the usual $\delta$-distribution is obtained in the configuration space:
\begin{equation}
\langle {\mathbf r} | \delta^D \rangle = \delta({\mathbf r}),
\end{equation}
and ${|\delta^D \rangle}$ is also denoted by the formal ket of zero relative coordinates ${|0_{\mathbf r} \rangle}$. From Eq.~\eqref{eq:scatt_state} the scattering state ${| \Psi_{\bf k_0}\rangle}$ is written in the zero-range potential approach as
\begin{equation}
| \Psi_{\bf k_0} \rangle = | \mathbf k_0 \rangle + \frac{T_D(E+i0^{+})}{E+i0^{+}-H_0}  |\delta^D\rangle.
\label{eq:ZRdiff}
\end{equation} 
Injecting Eq.~\eqref{eq:substitution} in the Lippmann-Schwinger equation~\eqref{eq:Lippmann-Schwinger} gives
\begin{equation}
| \Psi_{\bf k_0} \rangle = |\mathbf k_0\rangle + \frac{S_{\Psi_{\bf k_0}}}{E+i0^{+}-H_0}  |\delta^D\rangle .
\label{eq:def_Spsi}
\end{equation} 
Identifying Eq.~\eqref{eq:ZRdiff} and \eqref{eq:def_Spsi} shows that the source amplitude associated with a scattering state ${|\Psi_{\bf k_0}\rangle}$ is nothing but the on-shell transition matrix: ${S_{\Psi_{\bf k_0}} = T_D(E+i0^{+})}$. In configuration space, scattering states in Eq.~\eqref{eq:ZRdiff} or in Eq.~\eqref{eq:diff_r} are singular for vanishingly small interparticle distances. As shown in what follows, this singular behavior defines the Bethe-Peierls asymptotic condition. From  Eq.~\eqref{eq:ZRdiff} one finds that all the scattering states for a given dimension $D$ have the same singular behavior which reduces to
\begin{equation}
\langle {\mathbf r} | \Psi \rangle \operatornamewithlimits{=}_{{\mathbf r}\to 0}  A \times
\left\{ 
\begin{array}{ll}
\displaystyle  \left( \frac{1}{a_3} - \frac{1}{r}\right) + O(r) & \  {(D =3)} \\
\\
\displaystyle \ln\left(\frac{r}{a_2}\right)  + O(r) & \ {(D =2)} \\ 
\\
\displaystyle  \left(r - a_1 \right) + O(r) & \ {(D =1)} .
\end{array}
\right. 
\label{eq:contact_D}
\end{equation}
In Eq.~\eqref{eq:contact_D} $A$ is a function of the energy $E$ but does not depend on $r$. It is important to note that Eq.~\eqref{eq:contact_D} is obtained in the center-of-mass frame, meaning that the limit ${r\to 0}$ is taken at fixed value of the center-of-mass of the two colliding particles. Conversely, for a given dimension $D$, Eq.~\eqref{eq:contact_D} is the sufficient asymptotic condition which permits one to obtain the source amplitude ${S_{\Psi_{\mathbf k_0}}}$ of a scattering state ${|\Psi_{\mathbf k_0}\rangle}$ at energy $E$ in Eq.~\eqref{eq:def_Spsi}. The asymptotic condition in Eq.~\eqref{eq:contact_D} referred also as the ``contact condition'' or the Bethe-Peierls condition, remains unchanged for any linear combination of scattering states and can be thus used to find any eigenstate in the zero-range approximation for a system where an external potential is included within the free Hamiltonian $H_0$. For a positive energy $E$, the Lippmann Schwinger equation is
\begin{equation}
| \Psi \rangle = |\Psi^{(0)} \rangle + \frac{S_\Psi}{E+i0^{+}-H_0}  |\delta^D\rangle ,
\label{eq:formal}
\end{equation} 
where $|\Psi^{(0)} \rangle$ is the complementary solution \emph{i.e.}, it is an eigenstate of the free Hamiltonian ${H_0}$ at energy $E$ (for a negative energy  ${|\Psi^{(0)} \rangle=0}$ and  the ${+i0^+}$  prescription in Eq.~\eqref{eq:formal} is useless). The relation between the number $A$ in Eq.~\eqref{eq:contact_D} and the source amplitude ${S_\Psi}$ in Eq.~\eqref{eq:formal} is given by
\begin{equation}
A = \frac{2 \mu {\rm S}_{\Psi}}{\Omega_D \hbar^2} .
\end{equation}
The crucial point of this section is to show that the asymptotic condition in Eq.~\eqref{eq:contact_D} can be expressed in arbitrary representation. In what follows for convenience, the delta distribution is represented by the ${\epsilon \to 0}$ limit of a Gaussian weight: ${\delta^{D}({\mathbf r}) = \lim_{\epsilon  \to 0} \langle {\mathbf r} | \delta_\epsilon^D \rangle}$ and
\begin{equation}
\langle {\mathbf r} | \delta_\epsilon^D \rangle = \frac{1}{(2 \pi \epsilon^2)^{D/2}} \exp\left(-\frac{r^2}{\epsilon^2}\right) .
\label{eq:Gaussian}
\end{equation}
With this particular choice, in the momentum space the representation of the delta term has the same expression [denoted ${\chi_\epsilon(k)}$] for all dimensions:
\begin{equation}
\langle {\mathbf k}| \delta_\epsilon^D \rangle = \chi_\epsilon(k) = \exp \left(-\frac{k^2 \epsilon^2}{4} \right) .
\end{equation}
Matrix elements and states without index $\epsilon$ are considered in their zero-range limit which corresponds in this formalism to the limit ${\epsilon \to 0}$ (for example ${\lim_{\epsilon  \to 0} |\Psi_\epsilon \rangle = |\Psi \rangle}$. For the formulation of the zero-range approximation it is also useful to introduce the 'reference state' denoted ${| \phi_\epsilon^\Lambda \rangle}$ which results from the action of the two-body Green's function in free space on ${|\delta_\epsilon^D\rangle}$ at the negative energy ${E_\Lambda}$:
\begin{equation}
|\phi_\epsilon^\Lambda \rangle = \frac{1}{E_\Lambda-H_0} |\delta_\epsilon^D\rangle\quad   \mbox{where} \quad E_\Lambda = -\frac{\hbar^2\Lambda^2}{2\mu} <0 .
\label{eq:Green}
\end{equation}
In Eq.~\eqref{eq:Green} the parameter ${\Lambda}$ is chosen positive ${(\Lambda \in {\mathbb R}^{+})}$ according to the usual prescription in scattering theory. In the momentum representation:
\begin{equation}
\langle {\mathbf k}| \phi_\epsilon^\Lambda \rangle = - \frac{2\mu}{\hbar^2} \times \frac{\chi_\epsilon(k)}{k^2 + \Lambda^2 } .
\end{equation}
In the zero-range limit where ${\epsilon \to 0}$, in the configuration space the reference state has the same type of singularity as in Eq.~\eqref{eq:contact_D}:
\begin{equation}
\langle {\mathbf r} | \phi^\Lambda \rangle =   \frac{2\mu}{\Omega_D \hbar^2}
\left\{ 
\begin{array}{ll}
\displaystyle \left( - \frac{1}{r} + \Lambda \right) + O(r) &\ {(D =3)} \\
\\
\displaystyle \ln\left(\frac{\Lambda r e^\gamma}{2}\right)  + O(r) & \ {(D =2)} \\ 
\\
\displaystyle  \left(  r - \frac{1}{\Lambda}  \right) + O(r) &\ {(D =1)}.
\end{array}
\right. 
% - \frac{2\mu}{\hbar^2} \left( \frac{1}{r} - \Lambda \right)  + O(r2) .
\label{eq:phi_Lambda}
\end{equation}
The contact condition in Eq.~\eqref{eq:contact_D} can then be written in terms of the reference state as
\begin{equation}
\lim_{r \to 0} \lim_{\epsilon \to 0} \langle  {\mathbf  r}  | \Psi_\epsilon -  {\rm S}_{\Psi} \phi_\epsilon^\Lambda \rangle 
= \frac{{\rm S}_{\Psi}}{T_D(E_\Lambda)} .
\label{eq:BPepsilon}
\end{equation}
In the configuration space, the state ${| \Psi_\epsilon-{\rm S}_{\Psi}\phi_\epsilon^\Lambda \rangle}$ is a smooth function for ${(r,\epsilon)}$ close to ${(0,0)}$. It is thus possible to commute the ${r \to 0}$ and ${\epsilon \to 0}$ limits in Eq.~\eqref{eq:BPepsilon}:
\begin{equation}
\lim_{r \to 0} \lim_{\epsilon \to 0} \langle {\mathbf  r}|
\Psi_\epsilon - {\rm S}_{\Psi} \phi_\epsilon^\Lambda
\rangle = \lim_{\epsilon \to 0} \lim_{r \to 0}  \langle {\mathbf  r}|
\Psi_\epsilon - {\rm S}_{\Psi} \phi_\epsilon^\Lambda \rangle .
\end{equation}
For particles moving in the presence of an external potential, the source amplitude in configuration space depends on the center-of-mass coordinates of the interacting pair. Hence, in general situations the source amplitude is replaced by a state associated with the center-of-mass of the interacting pair:
\begin{equation}
S_\Psi \longrightarrow |{\rm S}_{\Psi}\rangle .
\end{equation}
Finally, the Bethe-Peierls condition can be written without specifying any representation as:
\begin{equation}
\lim_{\epsilon \to 0} \langle 0_{\mathbf r} | \Psi_\epsilon -  {\rm S}_{\Psi} \phi_\epsilon^\Lambda \rangle = \frac{|{\rm S}_{\Psi}\rangle}{T_D(E_\Lambda)} .
\label{eq:contact1}
\end{equation}
Equation \eqref{eq:contact1} is by construction, invariant in a change of ${\Lambda \in \mathbb{R}^{+}}$:  as is shown in the next section, one recovers the {\em so-called} ${\Lambda}$ freedom of the ${\Lambda}$-potential \cite{Ols02}. This freedom permits to simplify exact calculations without introducing any approximation, or also to improve approximate schemes \cite{Ols02,Pri04b,Pri06}. Moreover, the ${\epsilon \to 0}$ limit can be taken equivalently as follows,
\begin{equation}
\lim_{\epsilon \to 0} \langle \delta_\epsilon^D | \Psi_\epsilon  -  {\rm S}_{\Psi} \phi^\Lambda_\epsilon \rangle
=  \frac{|{\rm S}_{\Psi}\rangle}{T_D(E_\Lambda)} ,
\label{eq:contact2}
\end{equation}
or also by substituting the ket in the left hand side of Eq.~\eqref{eq:contact2} by its ${\epsilon \to 0}$ limit:
\begin{equation}
\lim_{\epsilon \to 0} \langle \delta_\epsilon^D | \Psi - {\rm S}_{\Psi} \phi^\Lambda \rangle = \frac{|{\rm S}_{\Psi}\rangle}{T_D(E_\Lambda)} .
\label{eq:contact3}
\end{equation}
The contact conditions in Eqs.~(\ref{eq:contact1})-(\ref{eq:contact3}) can be expressed in any desired representation by inserting a closure relation in the scalar product concerning the relative particle. For example in the momentum representation, one can insert the closure relation
\begin{equation}
\int \frac{d^D{k}}{(2\pi)^D}|{\mathbf k}\rangle \langle {\mathbf k} | = \mathbb{I} 
\end{equation}
at the right of the bra ${\langle \delta_\epsilon^D |}$ in Eq.~\eqref{eq:contact3} and this gives:
\begin{equation}
\lim_{\epsilon \to 0} \left( \int \frac{d^D{k}}{(2\pi)^D} \chi_\epsilon(k)
\langle {\mathbf k} | \Psi -  {\rm S}_{\Psi} \phi^\Lambda \rangle \right) 
=  \frac{|{\rm S}_{\Psi}\rangle}{T_D(E_\Lambda)} .
\label{eq:contact_k}
\end{equation}
In the next sections, it is shown that Eqs.~(\ref{eq:contact1})-(\ref{eq:contact3}) allow one to obtain in a simple way the standard integral equations of few-body problems in the zero-range limit and also to compute the induced scattering resonances and related scattering problems in the presence of a harmonic transverse confinement. As a conclusion of this part, it is interesting (and useful) to note that for ${D=3}$ or ${D=1}$, it is possible to eliminate the presence of the reference function in the contact condition by performing a specific limit on $\Lambda$. For ${D=3}$, in the  ${\Lambda \to 0}$ limit the contact condition can be written as:
\begin{equation}
\lim_{\epsilon \to 0}  \partial_\epsilon \left( \epsilon  \langle \delta_\epsilon^{D=3} | \Psi   \rangle  \right) =  \frac{\mu|{\rm S}_{\Psi}\rangle}{2\pi \hbar^2 a_3} .
\label{eq:contactD=3_reg}
\end{equation}
Equation~\eqref{eq:contactD=3_reg} can be written in the following alternative form:
\begin{equation}
\operatornamewithlimits{Reg}_{\epsilon \to 0} \langle \delta_\epsilon^{D=3} | \Psi \rangle = \frac{\mu|{\rm S}_{\Psi}\rangle}{2\pi \hbar^2 a_3} ,  
\label{eq:Regepsilon}
\end{equation}
where ${\displaystyle \operatornamewithlimits{Reg}_{\epsilon \to 0}}$ extracts the regular part of ${\langle \delta_\epsilon^{D=3}|\Psi \rangle}$ in the limit where ${\epsilon \to 0}$. For ${D=1}$, elimination of the reference function occurs in the limit where ${\Lambda}$ tends to ${\infty}$. One obtains:
\begin{equation}
\lim_{\epsilon \to 0}  \langle \delta_\epsilon^{D=1} | \Psi \rangle = - \frac{\mu a_1}{\hbar^2} | S_\Psi \rangle .
\label{eq:contactD=1_reg}
\end{equation}

\subsection{Link with the $s$ wave $\Lambda$ potential}

A zero-range $s$ wave pseudo potential was introduced by Fermi \cite{Fer36} in order to perform calculations in the first-order Born approximation. Thus, the Fermi pseudo potential cannot be  used for a non-perturbative approach (for large values of $a_3$, for example). In the modern formulation of the zero-range pseudo potential, the Bethe-Peierls condition is included in the Schr\"{o}dinger equation \emph{via} the zero-range pseudo potential itself \cite{Bre47,Bla52}. It has been shown in the configuration space that there exists in each dimension $D$ a family of zero-range potentials: the \emph{so-called} $\Lambda$ potentials \cite{Ols02}. This section, links the Bethe-Peierls approach and the $\Lambda$ potential in arbitrary representation. For this purpose, a family of operators ${{ R}^{\Lambda}_\epsilon}$ is introduced. They act on a state ${|\Psi_\epsilon\rangle}$ as: 
\begin{equation}
{ R}^{\Lambda}_\epsilon  |  \Psi_\epsilon \rangle =  | \delta_\epsilon^D \rangle  \lim_{\epsilon \to 0} \langle \delta_\epsilon^D 
| \Psi_\epsilon - {\rm S}_{\Psi} \phi^\Lambda_\epsilon \rangle .
\label{eq:RLambda_def}
\end{equation}
For a state verifying the contact condition in Eq.~\eqref{eq:contact2} one has:
\begin{equation}
{ R}^{(1/\tilde{a}_D)}_\epsilon  |  \Psi_\epsilon \rangle =  0   
\label{eq:RLambda_prop}
\end{equation}
where $\tilde{a}_D$ is related to the scattering length $a_D$ as
\begin{equation}
\tilde{a}_D= \left\{ 
\begin{array}{ll}
a_{D} &\ \mbox{for}\ D=3\ \mbox{or}\ D=1\\
a_2 e^\gamma/2 &\  \mbox{for}\ D=2
\end{array}
\right. .
\label{eq:aDtilde}
\end{equation}
These regularizing operator satisfy two other properties: $(i)$ for a regular state (${S_\Psi=0}$) thus ${{ R}^{\Lambda}_\epsilon|\Psi\rangle= |\delta_\epsilon^D \rangle  \langle 0_{\mathbf r} | \Psi\rangle}$ and $(ii)$ ${{ R}^{\Lambda}_\epsilon  |  \phi^\Lambda_\epsilon \rangle = 0}$.

Explicit expressions for ${{ R}^{\Lambda}_\epsilon}$ can be found without specifying any representation as follows:
\begin{equation}
{ R}^{\Lambda}_\epsilon  |  \Psi_\epsilon \rangle = | \delta_\epsilon^D \rangle \lim_{\epsilon \to 0} r^\Lambda_\epsilon 
\left[ \langle \delta_\epsilon^D | \Psi_\epsilon \rangle \right] ,
\label{eq:r_epsilon_Lambda_def}
\end{equation}
where ${r^\Lambda_\epsilon}$ is an operator defined in each dimension as: 
\begin{equation}
r^\Lambda_\epsilon \left[ \, \cdot \, \right]=
\left\{ 
\begin{array}{ll}
\displaystyle \left[ \left(\partial_\epsilon +\sqrt{\frac{\pi}{2}} \Lambda\right) \epsilon  \, \cdot  \, \right] &\  {(D=3)} \\
\\
\displaystyle \left[ \left( 1 + \frac{\epsilon}{2} \ln ( e^\gamma \Lambda^2 \epsilon^2/2 ) \partial_\epsilon \right)  \, \cdot  \, \right]   & \ {(D=2)} \\
\\
\displaystyle \left[ \left( 1 + \sqrt{\frac{\pi}{2}} \frac{1}{\Lambda}  \partial_\epsilon \right)  \, \cdot  \, \right]    & \ {(D=1)} . \\
\end{array}
\right.  \label{eq:r_lambda_epsilon_expr}
\end{equation}
Expressions of ${r^\Lambda_\epsilon}$ in Eq.~\eqref{eq:r_lambda_epsilon_expr} depend on the choice made for the short range functions 
${\langle \mathbf r |\delta_ \epsilon^D \rangle}$, given here by Eq.~\eqref{eq:Gaussian}. The source term in the stationary Schr\"{o}dinger equation can be expressed in terms of ${{ R}^{\Lambda}_\epsilon}$ by using Eqs.~(\ref{eq:contact1} and \ref{eq:RLambda_def}):
\begin{equation}
\left(H_0-E\right) |\Psi_\epsilon \rangle + T_D(E_\Lambda) { R}^{\Lambda}_\epsilon | \Psi_\epsilon \rangle = 0 .
\label{eq:H_Lambda}
\end{equation}
In the limit ${\epsilon \to 0}$, the pseudo potential in this equation coincides exactly with the ${\Lambda}$-potential introduced in Ref.~\cite{Ols02}, where the coupling constant ${g_\Lambda}$ is nothing but the transition matrix at energy ${E_\Lambda}$:
\begin{equation}
g_\Lambda =  T_D(E_\Lambda).
\end{equation}

\subsection{Context of resonant scattering} 

The zero-range approximation is especially interesting in the regime of parameters where the pairwise potential leads to a resonant scattering at low collisional energy. In this regime, the cross section proportional to ${|f_D|^2}$ has a large value at small relative momentum (${k_0 b_D \ll 1}$) and expressions in Eqs.~(\ref{eq:f3})-(\ref{eq:diff_r}) are good approximations. For ${D=3}$, this corresponds to the regime where the scattering length ${a_3}$ is large in absolute value as compared to the potential radius ${b_3}$ and the maximum is at ${k_0=0}$. This regime can be achieved by using the FR mechanism (see Ref.~\cite{Chi10} for a general review of FR in ultracold atoms). The FR involves the coupling between atoms (in the ``open channel'') and a molecular state (in the ``closed channel'') characterized by a size of the order of the potential range ${b_3}$. This two-channel description provides the expression of the scattering length (${a_3}$) as a function of the external magnetic field ${\mathcal B}$ in the vicinity of a given resonance located at ${\mathcal B=\mathcal B_0}$ which can be parameterized by the formula \cite{Moe95}
\begin{equation}
a_3 = a_{\rm bg} \left( 1 - \frac{\Delta \mathcal B}{\mathcal B - \mathcal B_0}
\right).
\label{eq:a3}
\end{equation}
In Eq.~\eqref{eq:a3}, ${\Delta \mathcal B}$ is the width of the magnetic FR and ${a_{\rm bg}}$ is the ``background'' scattering length \emph{i.e.}, the scattering length away from the FR. The magnetic width ${(\Delta \mathcal B)}$ can be also characterized by the ``width radius'' $R^\star$ defined by
\begin{equation}
R^\star = \frac{\hbar^2}{m a_{\rm bg} \delta \mu \Delta \mathcal B},
\end{equation}
where $\delta \mu$ is the difference of magnetic moment for an atomic pair in the open- and the closed-channel \cite{Pet04b,Chi10}. In the vicinity of a broad resonance, $R^\star$ is of the order of  (or smaller than) $b_3$ the 3D scattering amplitude can be approximated by Eq.~\eqref{eq:f3} in a large interval of momentum where ${|k_0| b_3 \ll 1}$. 

As shown in the next section, the resonant behavior in the low dimensional atomic waveguide can be reached by tuning the length $a_{\rm trans}$ of the transverse confinement and/or the 3D scattering length. For $D=1$ the resonant regime occurs in the limit where the 1D-scattering length ${a_{\rm 1}}$ is small with respect to the transverse length (\emph{i.e.}, ${|a_1|\ll b_1}$) and also for a vanishing relative momentum ${k_0 \to 0}$. In 2D, the maximum in ${|f_2|^2}$ occurs at the momentum ${k_0=2/(e^\gamma a_2)}$ (where  ${|f_2|=2/\pi}$) and the resonant regime is thus reached at low energy if and only if the 2D scattering length is large as compared to the transverse length (\emph{i.e.}, ${a_2 \gg b_2}$).

In the resonant regimes defined above, the probability of scattering for two colliding particles is large so that the pairwise short range potential affects the form of the wave function at interparticle distances which are large with respect to the potential radius ${b_D}$. More precisely, the scattering states in Eq.~\eqref{eq:scatt_state} can be decomposed in an incoming part ${|\phi_{\rm inc}\rangle=|\mathbf k_0\rangle}$ and a scattered part ${|\phi_{\rm scatt}\rangle}$. At large distance and in the resonant regime:
\begin{equation}
\left|\frac{\langle \mathbf r | \phi_{\rm scatt}\rangle}{\langle \mathbf r | \phi_{\rm inc}\rangle}\right| \operatornamewithlimits{=}_{r\gg b_D} 
\left\{
\begin{array}{ll} 
|a_3|/r & \ {(D=3)} \\
\sqrt{e^\gamma a_2}/\sqrt{\pi r} &\  {(D=2)}\\
1 & \ {(D=1)} .
\end{array}
\right.
\label{eq:res_large_r}
\end{equation}
For ${D=3}$ or ${D=2}$ this ratio is greater than or of the order of unity for ${b_D \ll r < |a_D|}$, where ${|a_D|}$ is arbitrarily large and is equal to unity for arbitrary large distance in one-dimensional systems (${D=1}$). The zero-range potential approach is a formalism which permits one to evaluate accurately the wave function in configurations where particles are outside the potential radius while configuration where two or more particles are inside the potential radius are not reliably described. Consequently, the Bethe-Peierls asymptotic approach is very well suited for studying systems in the resonant regime where the wavefunction is modified by the interaction at interparticle distances which are large as compared to the potential radius.

In the 3D space and for a positive scattering length, the on-shell transition matrix ${T_3}$ in Eq.~\eqref{eq:TD} has a real pole at negative energy ${E_2=-\hbar^2/(2\mu a_3^2)}$. In the resonant regime, this pole is associated with the existence of a shallow dimer which is thus very well described in the zero-range approach (the probability that the relative pair has a radius greater than the potential radius ${b_3}$ is ${\exp(-2b_3/a_3)\sim 1}$). For ${D=2}$, the pole is at ${E_2=-2 \hbar^2/(\mu a_2^2 \exp(2\gamma))}$, and in the resonant regime it is also the signature of the existence of a shallow dimer. For ${D=1}$, in the resonant regime (${a_1 \to 0}$) the approximation of the transition matrix in Eq.~\eqref{eq:TD} doesn't possess a negative low-energy pole and next-order terms in the energy expansion are needed to find the pole corresponding to the lowest bound state~\cite{Ber03}.

\section{Dimensional Reduction}
\label{sec:Dreduction}

In this section, the dimensional reduction issue is solved by using the zero-range approximation with the Bethe-Peierls condition expressed through Eq.~\eqref{eq:Regepsilon}. This section gives alternative derivations of the results given in Refs.~\cite{Ols98,Pet00,Pet01,Idz06}.
The problematic of the 3D$\to$1D (respectively 3D$\to$2D) dimensional reduction is as follows: two particles move freely in 1D (respectively 2D) while they are confined in the transverse direction by an atomic waveguide built from a 2D (respectively 1D) trapping potential. The collisional energy is such that at large interatomic distances, the particles are confined in the ground state of the trapping potential. Solving the issue amounts to finding the low energy transition matrix of this low dimensional quasi-2D (resp quasi-1D) scattering process as a function of the waveguide parameters and of the 3D scattering length $a_3$. 
In what follows only the case of harmonic trapping where the center-of-mass and relative motions decouple is considered. In the center-of-mass frame the source amplitude $S_\Psi$ is constant and the scattering state can be written as:
\begin{equation}
|\Psi \rangle = |\Psi^{(0)} \rangle + |\Psi^{\rm int} \rangle,
\label{eq:Psi_scatt}
\end{equation}
where ${|\Psi^{(0)} \rangle}$ is the incoming wave and  the ``interacting part'' is 
\begin{equation}
|\Psi^{\rm int} \rangle = \frac{{\rm S}_\Psi}{ E + i 0^{+}-H_0} |0_{\mathbf r} \rangle.
\label{eq:Psi_int}
\end{equation}
In Eq.~\eqref{eq:Psi_int}, the source amplitude ${{\rm S}_\Psi}$ is found from the Bethe-Peierls condition by solving the equation:
\begin{equation}
\langle 0_{\mathbf r} |\Psi^{(0)} \rangle + 
\operatornamewithlimits{Reg}_{\epsilon \to 0} \langle \delta_\epsilon^{\rm D=3}  | \Psi^{\rm int} \rangle =  \frac{\mu {\rm S}_{\Psi}}{2\pi\hbar^2 a_3} .
\label{eq:BPLowD}
\end{equation}
These quasi-1D and quasi-2D scattering problems have been initially solved by using the Bethe-Peierls method in the configuration space in Ref.~\cite{Ols98} and Refs.~\cite{Pet00,Pet01}. In this section, an alternative derivation is given in order to illustrate the formalism introduced in the previous section.
\subsection{Linear atomic waveguide}  
The 3D$\to$1D reduction problem, is considered here in the case where the two colliding atoms are confined in an isotropic two-dimensional harmonic trap while they move freely along the third direction ($z$). In this case, the problem can be solved in the center-of-mass frame where the noninteracting Hamiltonian is
\begin{equation}
H_0 = - \frac{\hbar^2}{2\mu} \partial_z^2 + {\mathcal H}_{\rm 2D},
\end{equation}
and ${\mathcal H}_{\rm 2D}$ is the Hamiltonian for the transverse motion:
\begin{equation}
{\mathcal H}_{\rm 2D}= - \frac{\hbar^2}{2\mu}  (\partial_x^2 + \partial_y^2 ) 
+ \frac{1}{2} \mu \omega_\perp^2 (x^2+y^2)  - \hbar\omega_\perp.
\label{eq:H2D}
\end{equation}
The linear atomic waveguide is characterized by the transverse length:
\begin{equation}
a_\perp=\sqrt{\frac{\hbar}{\mu \omega_\perp}}, 
\end{equation}
and the zero-range approximation of the ${D=3}$ pairwise potential is justified in the limit where ${a_\perp \gg b_3}$. In Eq.~\eqref{eq:H2D}, the zero-point energy has been subtracted. That way, the energy ${E}$ of the scattering states is measured with respect to the continuum threshold. In this quasi-1D scattering problem at energy ${E = \hbar^2k_0^2/2\mu}$, the incoming wave for the relative particle (state ${|\Psi^{(0)} \rangle}$) in Eq.~(\ref{eq:Psi_scatt}) has a momentum ${\mathbf k_0}$ along ${z}$ and is in the ground-state of the transverse 2D oscillator:
\begin{equation}
|\Psi^{(0)} \rangle = |n_x=0 \rangle |n_y=0 \rangle |k_z = k_0 \rangle , 
\end{equation}
where ${n_x}$ and ${|n_x\rangle}$ (respectively ${n_y}$ and ${|n_y\rangle}$) are the quantum number and eigenstate of the 1D harmonic oscillator of atomic frequency ${\omega_\perp}$ and mass $\mu$ along $x$ (respectively $y$). The system is quasi-1D for collisional energies such that:
\begin{equation}
E < E_{\rm trans}^{\rm 1D} ,
\label{eq:monomode_1D}
\end{equation}
where ${E_{\rm trans}^{1D}=2 \hbar \omega_\perp}$. Equation \eqref{eq:monomode_1D} defines the monomode regime of the atomic waveguide such that outgoing particles are trapped at large relative distances in the ground state of the transverse trap. The contribution of the incoming wave in Eq.~\eqref{eq:BPLowD} is:
\begin{equation}
\langle 0_{\mathbf r} |\Psi^{(0)} \rangle =  \phi_0(0)^2 = \frac{1}{\sqrt{\pi a_\perp}} ,
\end{equation}
where $\phi_n(x) = \langle x | n \rangle$ is given by:
\begin{equation}
\phi_n(x) = \frac{\exp \left(\frac{-x^2}{2a_\perp^2} \right)}{\pi^{1/4} \sqrt{a_\perp}}  H_n\left(\frac{x}{a_\perp}\right).
\label{eq:phi_n}
\end{equation}
In the calculation of ${\langle \delta_\epsilon^{\rm D=3}|\Psi^{\rm int}\rangle}$, the quantum numbers of the noninteracting Hamiltonian  are introduced directly by inserting the closure relation:
\begin{equation} 
\int_{-\infty}^{+\infty}\frac{dk_z}{2\pi}| k_z \rangle \langle k_z | 
\sum_{n_x=0}^\infty |n_x\rangle \langle n_x| 
\sum_{n_y=0}^{\infty} |n_y\rangle \langle n_y|= \mathbb{I}.
\end{equation}
Only even values of ${n_x}$ and ${n_y}$ give a non vanishing contribution, and one obtains
\begin{multline}
\langle \delta_\epsilon^{\rm D=3}| \Psi^{\rm int} \rangle = \frac{{\rm S}_\Psi}{\hbar \omega_\perp} 
\int_{-\infty}^{+\infty} \frac{dk_z}{2\pi} \sum_{p,q=0}^\infty \exp\left(-\frac{k_z^2 \epsilon^2}{4}\right) \times \\
\frac{\langle \delta_\epsilon^{\rm D=1} | 2p \rangle \langle \delta_\epsilon^{\rm D=1} | 2q \rangle 
\phi_{2p}^*(0) \phi_{2q}^*(0)}{\tau-\frac{k_z^2a_\perp^2}{4}-p-q},
\label{eq:delta_Psi_int_2D}
\end{multline}
where the dimensionless energy variable $\tau$ is defined by
\begin{equation}
\tau=\frac{E}{2\hbar\omega_\perp}+i0^{+}.
\label{eq:positiveE}
\end{equation}
The behavior of ${\langle \delta_\epsilon^{\rm D=3}|\Psi^{\rm int} \rangle}$ in the limit where $\epsilon$ vanishes is more easily obtained in the domain of negative collisional energies ${(\tau <0)}$ than in the domain of positive collisional energies ${E>0}$ and are related the each to the other by analyticity. For ${\tau<0}$ one can use the identity
\begin{equation}
\frac{1}{\frac{k_z^2 a_\perp^2}{4} + n + p - \tau } = \int_0^\infty du\, e^{-u \left( \frac{k_z^2 a_\perp^2}{4}+n+p-\tau \right)},
\label{eq:trick}
\end{equation}
which allows one to decouple the discrete summations from the integration over ${k_z}$ in Eq.~\eqref{eq:delta_Psi_int_2D}. From Eq.~\eqref{eq:phi_n}, one can deduce the following limit:
\begin{equation}
\lim_{\epsilon \to 0} \sum_{p=0}^\infty \langle \delta_\epsilon^{\rm D=1} | 2p \rangle  \phi_{2p}^{*}(0) e^{-pu} 
= \frac{|\phi_0(0)|^2}{\sqrt{1-e^{-u}}}.
\label{eq:sum_epsilon}
\end{equation}
Using Eq.~\eqref{eq:sum_epsilon} and integrating over $k_z$ in Eq.~\eqref{eq:delta_Psi_int_2D} gives:
\begin{equation}
\langle \delta_\epsilon^{\rm D=3}  | \Psi^{\rm int} \rangle \operatornamewithlimits{=}_{\epsilon \to 0} 
\frac{-\mu {\rm S}_\Psi}{2\pi^{3/2} \hbar^2 a_\perp}  
\int_0^{\infty} \frac{du}{\sqrt{u+\frac{\epsilon^2}{a_\perp^2}}} \frac{e^{u\tau}}{1-e^{-u}}.
\end{equation}
In the limit where ${\epsilon \to 0}$, this expression diverges as ${1/\epsilon}$ and its regular part coincides with a Hadamard's finite part \cite{Gel64,Sch78}:
\begin{equation}
\operatornamewithlimits{Reg}_{\epsilon \to 0} \langle \delta_\epsilon^{D=3}|\Psi^{\rm int} \rangle 
= \frac{ \mu {\rm S}_\Psi}{2\pi^{3/2} \hbar^2 a_\perp} \operatornamewithlimits{P.f.} \int_0^\infty \frac{du}{\sqrt{u}} \frac{e^{u \tau}}{1-e^{-u}}.
\label{eq:PF}
\end{equation}
One recognizes in Eq.~\eqref{eq:PF} an expression similar to the integral representation of the Hurwitz zeta function ${\zeta(s,z)}$ defined for ${\Re(s)>1}$ and ${0<z<1}$ by \cite{Gra94}:
\begin{equation}
\zeta(s,z) = \frac{1}{\Gamma(s)} \int_0^\infty dt \frac{t^{s-1} e^{-t z}}{1-e^{-t}} .
\label{eq:zetaH}
\end{equation}
However, in Eq.~\eqref{eq:zetaH} the integral diverges for ${s=1/2}$. The Hadamard's finite part in Eq.~\eqref{eq:PF} permits one to achieve a meromorphic continuation in the variable ${s}$ of Eq.~\eqref{eq:zetaH} and to identify the function ${\zeta(1/2,z)}$ \cite{Meromorphic}. Hence,
\begin{equation}
{\rm S}_\Psi = \frac{2\pi^{1/2}\hbar^2}{\mu \left[\zeta\left(\frac{1}{2},-\tau\right) + \frac{a_\perp}{a_3}\right]}.
\label{eq:S_Psi_zeta}
\end{equation}
For positive energies and in the monomode regime (${0<E<2\hbar\omega_\perp}$), the scattering state written in the configuration representation in Eq.~\eqref{eq:Psi_scatt}  has a non evanescent contribution at large relative length ${z \gg a_\perp}$, in the subspace of the transverse ground-state only. This property allows one to identify a quasi-1D scattering process with:
\begin{multline}
\langle {z, n_x=0,n_y=0} | \Psi \rangle \operatornamewithlimits{=}_{|z| \gg a_\perp} \exp(ik_0z) - \\ 
\frac{i \mu {\rm S}_\Psi}{\sqrt{\pi} \hbar^2 a_\perp k_0} \exp(ik_0 |z|) ,
\end{multline}
and ${a_\perp}$ plays the role of a 1D-potential radius (\emph{i.e.}, ${b_1 \sim a_{\perp}}$). The low-energy scattering amplitude ${f^{\rm 3D \to 1D}}$ of this quasi-1D system is thus:
\begin{equation}
f^{\rm 3D \to 1D}(k_0) = - \frac{i \mu {\rm S}_\Psi}{\sqrt{\pi} \hbar^2 a_\perp k_0}.
\label{eq:f3D>1D}
\end{equation}
In the ${|k_z\rangle}$-representation, the scattering states read
\begin{multline}
\langle {k_z, n_x=0,n_y=0} | \Psi \rangle = (2\pi) \delta(k_z-k_0) +\\ 
\frac{{\rm S}_\Psi}{\sqrt{\pi} a_\perp \left(E+i0^{+} -\frac{\hbar^2k_z^2}{2\mu}\right)} ,
\end{multline}
so that the quasi-1D scattering transition matrix ${T^{{\rm 3D} \to {\rm 1D}}}$ can be expressed as
\begin{equation}
T^{{\rm 3D} \to {\rm 1D}} = \frac{{\rm S}_\Psi}{\sqrt{\pi} a_\perp}.
\label{eq:T3D>1D}
\end{equation}
Finally, from Eqs.~(\ref{eq:S_Psi_zeta}) and (\ref{eq:T3D>1D}) one obtains:
\begin{equation}
T^{{\rm 3D} \to {\rm 1D}} = \frac{2 \hbar^2}{\mu a_\perp} \times \frac{1}{\zeta(\frac{1}{2},-\tau) + \frac{a_\perp}{a_3}}.
\end{equation}
In the low energy limit (${k_0 a_\perp \ll 1}$),
\begin{equation}
\zeta\left(\frac{1}{2},-\tau \right) \operatornamewithlimits{\simeq}_{\tau\to 0} \frac{1}{\sqrt{-\tau}} + \zeta\left(\frac{1}{2}\right)
\end{equation}
and for positive energies ${\sqrt{-\tau}=-ik_0 a_\perp/2}$. From Eq.~\eqref{eq:TD}, one can then identify a 1D-scattering length which is a function of the transverse length ${a_\perp}$ and of the 3D scattering length ${a_3}$ \cite{Ber03}:
\begin{equation}
a_1 = - \frac{a_\perp}{2} \left[ \frac{a_\perp}{a_3}+ \zeta\left(\frac{1}{2} \right)\right] .
\end{equation}
\subsection{Planar atomic waveguide}
In the 3D$\to$2D reduction problem, the two colliding particles are confined in a planar harmonic waveguide of frequency ${\omega_z}$ along the $z$ direction while they move freely in the two other directions. The noninteracting Hamiltonian in the center-of-mass frame reads:
\begin{equation}
H_0 =- \frac{\hbar^2}{2\mu} (\partial_x^2+\partial_y^2) + {\mathcal H}_{z},
\end{equation} 
with, the 1D-harmonic trap along $z$,
\begin{equation}
\mathcal H_{z} = - \frac{\hbar^2}{2\mu} \partial^2_z + \frac{\mu}{2} \omega_z^2 z^2 -\frac{\hbar \omega_z}{2} .
\label{eq:Hplanar}
\end{equation}
In Eq.~\eqref{eq:Hplanar} the zero point energy of the transverse trap has been subtracted, so that similarly to the previous {3D$\to$1D} reduction problem, the  energy of an eigenstate is measured with respect to the continuum threshold and coincides with the collisional energy. In what follows, ${|n_z\rangle}$ is the eigenstate of quantum number ${n_z}$ for the 1D harmonic oscillator of frequency ${\omega_z}$ and mass $\mu$. The subsequent derivation supposes that the 3D zero-range approximation is justified, \emph{i.e.}, that the atomic waveguide is such that:
\begin{equation}
a_z \gg b_3,
\end{equation}
where ${a_z}$ is the characteristic length of the atomic waveguide:
\begin{equation}
a_z=\sqrt{\frac{\hbar}{\mu\omega_z}}.
\end{equation}
In the quasi-2D scattering problem the incoming state is:
\begin{equation}
|\Psi^{(0)}\rangle=|{\mathbf k}_{\rm 2D} = {\mathbf k}_0 \rangle | n_z = 0 \rangle  ,
\end{equation}  
where ${\mathbf k}_0$ is the 2D-wave relative wavevector of the incoming wave and ${{\mathbf k}_{\rm 2D} = k_x \hat{\mathbf e}_{x} + k_y \hat{\mathbf e}_{y}}$ is the 2D collisional relative momentum. The collisional energy is ${E=\hbar^2k_0^2/2\mu}$ and the monomode regime condition is given by:
\begin{equation}
E < E_{\rm trans}^{\rm 2D}
\label{eq:monomode_2D}
\end{equation}
where ${E_{\rm trans}^{\rm 2D}=\hbar \omega_z}$. Equation~\eqref{eq:monomode_2D} ensures that the system is quasi-2D; \emph{i.e.}, the colliding particles are asymptotically trapped in the ground state of the transverse trap. The behavior of ${\langle \delta_\epsilon^{\rm D=3}|\Psi^{\rm int} \rangle}$ in the limit where $\epsilon \to 0$ is obtained by using the same techniques as for the 3D$\to$1D reduction problem. The quantum numbers of the free Hamiltonian are introduced by insertion of the closure relation: 
\begin{equation} 
\int\frac{d^2{k}_{\rm 2D}}{(2\pi)^2}  | {\mathbf k}_{\rm 2D} \rangle \langle {\mathbf k}_{\rm 2D} |   
\sum_{n_z=0}^\infty | n_z \rangle \langle n_z | = \mathbb{I} .
\end{equation}
In the 3D Bethe-Peierls asymptotic condition, only even values of ${n_z}$ contributes and one obtains:
\begin{multline}
\langle \delta_\epsilon^{\rm D=3} | \Psi^{\rm int} \rangle = \frac{\mu{\rm S_\Psi}a_z^2}{4\pi\hbar^2} \int_0^\infty k_{\rm 2D} dk_{\rm 2D}   \\  
\times 
\sum_{p=0}^\infty 
\exp\left(-\frac{k_{\rm 2D}^2 \epsilon^2}{4}\right)
\frac{\langle \delta_\epsilon^{\rm D=1}  | 2p \rangle \phi_{2p}^*(0)}{\tau-\frac{k_{\rm 2D}^2a_z^2}{4}-p},
\label{eq:Zepsilon}
\end{multline}
where the dimensionless energy variable $\tau$ is defined by
\begin{equation}
\tau= \frac{E}{2\hbar\omega_z} +i0^{+}.
\label{eq:positiveE2}
\end{equation}
The discrete summation in Eq.~\eqref{eq:Zepsilon} is performed in the domain ${\tau < 0}$ by using the transformation of Eq.~\eqref{eq:trick} together with the identity of Eq.~\eqref{eq:sum_epsilon}. After integration over ${k_{\rm 2D}}$, the regular part of ${\langle \delta_\epsilon^3 | \Psi^{\rm int} \rangle}$ can be expressed in the limit where ${\epsilon \to 0}$ as
\begin{equation}
\operatornamewithlimits{Reg}_{\epsilon \to 0} \langle \delta_\epsilon^3 | \Psi^{\rm int} \rangle =  - \frac{\mu{\rm S}_\Psi}{2\pi\hbar^2} |\phi_0(0)|^2  J(\tau) ,
\end{equation}
where the function $J(\tau)$  is  defined in the domain ${\tau<0}$ by: 
\begin{equation}
J(\tau) =  \mbox{P.f.}\! \int_0^\infty \!\! \frac{du}{u} \frac{\exp(\tau u)}{\sqrt{1-\exp(-u)}}.
\label{eq:J}
\end{equation}
From Eq.~\eqref{eq:BPLowD}, one finally obtains:
\begin{equation}
{\rm S}_\Psi = \frac{2\pi \hbar^{2} \phi_0(0)}{\mu \left[\frac{1}{a_3} + |\phi_0(0)|^{2} J(\tau) \right]}.
\end{equation}
Projection of the wavefunction on the ground-state of the 1D transverse harmonic oscillator ${\langle {\mathbf k}_{\rm 2D}, n_z = 0 | \Psi \rangle}$ gives the quasi-2D transition matrix:
\begin{equation}
T^{{\rm 3D} \to {\rm 2D}} = - \frac{\pi \hbar^{2}}{\mu \left[\frac{\sqrt{\pi} a_z}{2 a_3} +   \frac{J(\tau)}{2}   \right] }.
\end{equation}
In the domain of positive energy Eq. \eqref{eq:positiveE2} is continued analytically from the identity
\begin{equation}
J(\tau)=\ln\left(-\frac{B}{2\pi \tau}\right)+\sum_{n=1}^\infty \ln\left(\frac{n}{n-\tau}\right) \frac{(2n-1)!!}{(2n)!!},
\label{eq:Jtau-series}
\end{equation}
where ${B \simeq  0.9049}$ is defined by \cite{Pri08b}:
\begin{equation}
\ln\left(\frac{B e^\gamma}{2\pi}\right) = 
\int_0^\infty \!\! du\, \left( \frac{u^{-1}}{\sqrt{1-e^{-u}}}-\frac{1}{u^{3/2}}-\frac{1}{1+u} \right) ,
\label{eq:B}
\end{equation}
and $\gamma=0.5772\dots$ is the Euler's constant. The 2D low energy condition is given by:
\begin{equation}
 |E| \ll \hbar \omega_z 
\label{eq:LowE_2D}
\end{equation}
and ${a_z}$ plays the role of a 2D-potential radius, \emph{i.e.}, ${b_2 \sim a_z}$. Only the first logarithmic term in Eq.~\eqref{eq:Jtau-series} contributes at the lowest order in the limit of Eq.~\eqref{eq:LowE_2D}, and one can identify the resulting expression of the quasi-2D transition matrix ${T^{{\rm 3D} \to {\rm 2D}}}$ with the 2D low energy transition matrix ${T_2}$ in Eq.~\eqref{eq:TD}. The 2D scattering length ${a_2}$ of this quasi-2D scattering problem can thus be expressed in terms of $a_z$ and of the 3D scattering length ${a_3}$ \cite{Pet01,Idz06,Pri06} as:
\begin{equation} 
a_2  = a_z e^{-\gamma} \sqrt{\frac{2\pi}{B}} 
\exp\left(-\frac{a_z\sqrt{\pi}}{2 a_3}\right) . \label{eq:a2D}
\end{equation}
It is interesting to discuss a little bit further the condition for having a low energy resonant behavior in this quasi-2D system. The 2D cross-'section' (this is a length) is proportional to ${|T^{{\rm 3D} \to {\rm 2D}}(k_0)|^2/k_0}$ and a maximum occurs in the low-energy regime Eq.~\eqref{eq:LowE_2D} for:
\begin{equation}
\frac{a_z}{a_3} = \frac{1}{\sqrt{\pi}} \ln \left( \frac{\pi E}{B\hbar \omega_z }\right).
\label{eq:2DCIR}
\end{equation}
The right hand side of Eq.~\eqref{eq:2DCIR} is large and negative which shows that the resonance occurs for sufficiently large values of ${|a_3|}$ with respect to ${a_z}$  in the domain of negative 3D scattering lengths ${(a_3<0)}$. At fixed value of the collisional energy $E$ and of the 3D scattering length, the resonance can be reached by tuning the harmonic frequency of the atomic waveguide: this is the {\em so-called} quasi-2D confinement induced resonance first found in Ref.~\cite{Pet00}.

\section{Formalism for $N$-body systems}

\label{sec:Nbody}

\subsection{${\mathbf k}$ representation}

In this section the Bethe-Peierls method is applied to few- and many-body systems in the ${\mathbf k}$ representation. The system is composed of $N$ particles of respective masses ${m_1, m_2, \dots m_N}$ and of momenta ${{\mathbf k}_1, {\mathbf k}_2, \dots {\mathbf k_N}}$. For convenience the set of momenta is denoted by the short-hand notation $\{ {\mathbf k} \}$. The relative and total momenta of a given pair of particles $(ij)$,  are denoted by
\begin{equation}
{\boldsymbol k}_{ij} = \frac{m_j {\mathbf k}_i- m_i {\mathbf k}_j}{m_i + m_j} \ ,\ {\boldsymbol K}_{ij} = {\mathbf k}_i+{\mathbf k}_j ,
\end{equation}
and the  reduced mass of the pair $(ij)$ is denoted ${\mu_{ij}}$:
\begin{equation}
\mu_{i,j}=\frac{m_i m_j}{m_i+m_j}
\end{equation}
It is also convenient to introduce the set of momenta denoted $\{ {\boldsymbol \xi}^{ij} \}$ with ${\{ {\boldsymbol \xi}^{ij}  \}={\boldsymbol \xi}^{ij}_1, {\boldsymbol \xi}^{ij}_2 \dots {\boldsymbol \xi}^{ij}_N}$ where:
\begin{equation}
{\boldsymbol \xi}_n^{ij} = \left\{
\begin{array}{ll}
\displaystyle {\boldsymbol K}_{ij} \sqrt{\frac{m_i}{2(m_i+m_j)}}  & \mbox{for}\, n = i  \\
\displaystyle {\boldsymbol K}_{ij} \sqrt{\frac{m_j}{2(m_i+m_j)}}  & \mbox{for}\, n = j  \\
{\mathbf k}_n   & \mbox{otherwise} \\
\end{array}
\right. .
\end{equation}
This last notation permits one to isolate in the kinetic energy the contribution of the relative particle $(ij)$ from the ${N-1}$ other momenta:
\begin{equation}
\sum_{n=1}^N \frac{{\mathbf k}_n^2}{m_n} = \frac{{\boldsymbol k}_{ij}^2}{\mu_{i,j}} + \sum_{n=1}^{N} \frac{\left({\xi}_n^{ij} \right)^2}{m_n} .
\end{equation}
A non vanishing source amplitude ${|{\rm S}^{i \leftrightharpoons j}_{\Psi}\rangle}$ is associated with any pair of particles interacting via the short range pairwise potential in the many-body state ${|\Psi_{\epsilon}\rangle}$. In the case where the system is composed of particles with spin, in what follows ${|\Psi_{\epsilon}\rangle}$ denotes the projection over a given spin configuration of the many-body state including the symmetry imposed by the quantum statistics. The free Hamiltonian of the system (which may include an external potential) is denoted  ${H_0}$, and the ${N}$-body stationary state at energy $E$ verifies:
\begin{equation}
( H_0 - E ) | \Psi_\epsilon \rangle =-\sum_{i<j} 
|(ij):\delta_\epsilon^D \rangle |{\rm S}^{i \leftrightharpoons j}_{\Psi} \rangle .
\label{eq:many_body}
\end{equation}
In Eq.~\eqref{eq:many_body}, ${|(ij):\delta_\epsilon^D \rangle}$ is the  state ${|\delta_\epsilon^D \rangle}$ for the relative particle ${(ij)}$ of momentum ${{\boldsymbol k}_{ij}}$, and the source amplitude ${|{\rm S}^{i \leftrightharpoons j}_{\Psi}\rangle}$ is a state for the ${N-1}$ other particles of momenta ${\{{\boldsymbol \xi}^{ij}\}}$. If the pair ${(ij)}$ does not interact then
${|{\rm S}^{i \leftrightharpoons j}_{\Psi}\rangle=0}$,  otherwise it satisfies the Bethe-Peierls contact condition:
\begin{equation}
\lim_{\epsilon \to 0} \langle (ij):\delta_\epsilon^D | \left( | \Psi \rangle -  |(ij):\phi^\Lambda \rangle  |{\rm S}^{i \leftrightharpoons j}_\Psi\rangle  \right)
%\delta_\epsilon^D \rangle  \langle  \{ {\boldsymbol \xi_l^{ij}} \} |{\rm S}_{\Psi}^{i \leftrightharpoons j}_\Psi\rangle }{{\boldsymbol k}_{ij}^2 + \Lambda^2}  
=\frac{ | {\rm S}^{i \leftrightharpoons j}_{\Psi}\rangle}{T_D(E_\Lambda) }  , 
\label{eq:contact_Nbody}
\end{equation}
where the reference state  ${|(ij):\phi^\Lambda_\epsilon \rangle}$ is defined in the ${\mathbf k}$ representation as:
\begin{equation}
\langle {\boldsymbol k}_{ij}|(ij):\phi^\Lambda_\epsilon \rangle = - \frac{2\mu_{i,j}}{\hbar^2 } \times \frac{\chi_\epsilon(k_{ij})}{k_{ij}^2+ \Lambda^2} \end{equation}
In what follows Eqs.~(\ref{eq:many_body}) and (\ref{eq:contact_Nbody}) are used in the case where there is no external potential and ${|\Psi^{(0)}\rangle}$ denotes the complementary solution of Eq.~\eqref{eq:many_body} which satisfies the boundary conditions of the problem considered (for example an incoming plane-wave for a scattering problem). The many-body wavefunction can then be written as
\begin{equation}
\langle \{ {\mathbf k} \} | \Psi_\epsilon \rangle
=\langle \{ {\mathbf k} \} | \Psi^{(0)} \rangle +
{\mathcal G}_E(\{{\mathbf k}\})
\sum_{i<j} \chi_\epsilon(k_{ij}) \langle \{ {\boldsymbol \xi^{ij}} \} | {\rm S}^{i \leftrightharpoons j}_{\Psi} \rangle
\label{eq:NbodyWF}
\end{equation}
where ${{\mathcal G}_E(\{{\mathbf k}\})}$ is the $N$-body Green's function in the ${\mathbf k}$ representation:
\begin{equation}
{\mathcal G}_E(\{{\mathbf k}\}) = \frac{1}{E + i0^{+} - \sum_{n=1}^N \frac{\hbar^2 k_n^2}{2m_n} }
\end{equation}
For each interacting pair, the contact condition Eq.~\eqref{eq:contact_Nbody} gives in the ${\mathbf k}$ representation an integral equation in terms of the source amplitudes and of the complementary solution ${|\Psi^{(0)}\rangle}$. Interestingly, one can extract the contribution of the interacting pair ${(ij)}$ in the contact condition Eq.~\eqref{eq:contact_Nbody} without any integration by using the fact that this equation is invariant in a change of ${\Lambda\in \mathbb{R}^{+}}$ (\emph{i.e.}, by using the ${\Lambda}$ freedom). To this end, one expresses the Green's function in Eq.~\eqref{eq:NbodyWF} as:
\begin{equation}
{\mathcal G}_E(\{{\mathbf k}\}) = \frac{1}{E_{\rm col}^{(ij)} + i0^{+} -\frac{\hbar^2 k_{ij}^2}{2 \mu_{i,j}}} ,
\end{equation}
where ${E_{\rm col}^{(ij)}}$ is the collisional energy of the pair ${(ij)}$ defined by:
\begin{equation}
E^{\rm (ij)}_{\rm col} = E - \sum_{n=1}^{N} \frac{\hbar^2 \left( {\xi}_n^{ij} \right)^2 }{2m_n} .
\end{equation}
It is the energy of the pair ${(ij)}$ in its own center-of-mass frame while the other particles do not interact for a given set of momenta $\{{\mathbf k} \}$ and a total energy $E$. 

For a negative energy (${E<0}$) in which case ${| \Psi^{(0)}  \rangle=0}$, {\it without loss of generality} one can make the particular choice ${\Lambda^2=-2 \mu_{i,j} E^{\rm (ij)}_{\rm col}/\hbar^2}$ in Eq.~\eqref{eq:contact_Nbody}. That way, the term involving the reference state $\phi^\Lambda$ exactly cancels with the term associated with the source amplitude of the pair $(ij)$ (${|{\rm S}^{i \leftrightharpoons j}_{\Psi} \rangle}$). By analytical continuation, the same integral equation holds for ${E>0}$ (in which case  ${| \Psi^{(0)}  \rangle \ne 0}$). Finally, assuming that the ${\epsilon \to 0}$ limit is well defined, for each interacting pair one obtains the following integral equation:
\begin{multline}
 \operatornamewithlimits{\sum_{n<p}}_{(n,p)\ne(i,j)} 
\int \frac{d^D{k}_{ij}}{(2\pi)^D}  {\mathcal G}_E(\{{\mathbf k}\}) \langle  
\{ {\boldsymbol \xi^{np}} \} |{\rm S}^{n \leftrightharpoons p}_\Psi\rangle \\
= \frac{\langle \{ {\boldsymbol \xi^{ij}} \} | {\rm S}^{i \leftrightharpoons j}_{\Psi} \rangle}{T_D( E^{\rm (ij)}_{\rm col}+i0^{+})} 
- \int \frac{d^D{k}_{ij}}{(2\pi)^D} \langle \{{\mathbf k} \} | \Psi^{(0)}  \rangle .
\label{eq:integral_condition}
\end{multline}
One has to notice that the integrals in the first line of Eq.~\eqref{eq:integral_condition} are performed with the constraint that ${\{\boldsymbol \xi^{ij}\}}$ is held fixed (but ${\{\boldsymbol \xi^{np}\}}$ is not !).

\subsection{Second quantization}

In this section only fermionic particles of the same mass $m$ and two spin-components are considered. Generalization to other systems gives similar equations. The creation and annihilation operator of an atomic wave of momentum ${\mathbf k}$ for a spin $\sigma$ are denoted ${a_{{\mathbf k},\sigma}^\dagger}$ and ${a_{{\mathbf k},\sigma}}$ with the standard anti-commutation rule:
$\{a_{{\mathbf k},\sigma},a_{{\mathbf k},\sigma'}^\dagger \}=(2\pi)^3\delta(\mathbf k - \mathbf k')\delta_{\sigma \sigma'}$
and $\{a_{{\mathbf k},\sigma},a_{{\mathbf k},\sigma'} \}=0.$

The general expression of the ${\Lambda}$-potential is a simple way to write the Hamiltonian in its second quantized form. Using the definition of the $\Lambda$ potential for finite values of $\epsilon$ in Eq.~\eqref{eq:H_Lambda}, the Hamiltonian can be expressed as:
\begin{multline}
H_\epsilon = \int \frac{d^D{k}}{(2\pi)^D} \sum_\sigma \frac{\hbar^2 k^2}{2m} a_{{\mathbf k},\sigma}^\dagger a_{{\mathbf k},\sigma}
+\frac{T_D(E_\Lambda)}{2} \int \frac{d^D{K}}{(2\pi)^{D}}  \\
\times  \int \frac{d^D{k}'}{(2\pi)^{D}} \chi_\epsilon(k') 
a_{\frac{1}{2}{\mathbf K}-{\mathbf k}',\uparrow}^\dagger a_{\frac{1}{2}{\mathbf K}+{\mathbf k}',\downarrow}^\dagger\\
\times \lim_{\epsilon\to 0} r^\Lambda_\epsilon 
\left[ \int \frac{d^D{k}}{(2\pi)^{D}}  \chi_\epsilon(k) a_{\frac{1}{2}{\mathbf K}-{\mathbf k},\downarrow} 
a_{\frac{1}{2}{\mathbf K}+{\mathbf k},\uparrow} \ \cdot \ \right] ,
\end{multline}
where the dot ($\cdot$) after the annihilation operator ${a_{\frac{1}{2}{\mathbf K}+{\mathbf k},\uparrow}}$ reminds that the limit ${\epsilon \to 0}$ depends on the many-body state on which ${H_\epsilon}$ applies. 

In the hypothesis where the few- or many-body state ${|\Psi_\epsilon\rangle}$ is well defined in the zero-range limit (where only the Bethe-Peierls contact condition is used), the mean energy of the system ${\langle \Psi_\epsilon | H_\epsilon | \Psi_\epsilon \rangle}$ is a regular function of $\epsilon$ and the energy theorem \cite{Ols03,Tan08a,Tan08b,Com09,Wer10} follows from the identity:
\begin{equation}
\lim_{\epsilon \to 0} r^{(1/\tilde{a}_D)}_\epsilon \left[ \langle \Psi_\epsilon | H_\epsilon | \Psi_\epsilon \rangle \right] = \langle H \rangle .
\end{equation}
The state ${|\Psi_\epsilon\rangle}$ verifies the contact condition for each interacting pair thus  using the property in Eq.~\eqref{eq:RLambda_prop}, action of the operator ${\lim_{\epsilon \to 0} r^{(1/\tilde{a}_D)}_\epsilon \left[ \cdot \right]}$ on the interacting term in ${\langle \Psi_\epsilon | H_\epsilon | \Psi_\epsilon \rangle}$ gives exactly zero and:
\begin{equation}
\langle H \rangle = \lim_{\epsilon \to 0} r^{(1/\tilde{a}_D)}_\epsilon \left[ \int \frac{d^D{k}}{(2\pi)^D} 
\sum_\sigma \frac{\hbar^2 k^2}{2m} \langle a_{{\mathbf k},\sigma}^\dagger a_{{\mathbf k},\sigma}\rangle_\epsilon \right] .
\label{eq:E_Th}
\end{equation}
The energy of the system can be thus expressed in terms of the one-body density:
\begin{equation}
n_{{\mathbf k},\sigma} = \lim_{\epsilon \to 0}  \langle a_{{\mathbf k},\sigma}^\dagger a_{{\mathbf k},\sigma}\rangle_\epsilon .
\end{equation}
In the limit where ${\epsilon \to 0}$, one obtains from Eq.~\eqref{eq:NbodyWF} the high-momentum behavior of $n_{{\mathbf k},\sigma}$:
\begin{equation}
n_{{\mathbf k},\sigma} \operatornamewithlimits{=}_{k \to \infty}  \frac{\rm S}{k^4} + O\left(\frac{1}{k^6}\right),
\end{equation}
where ${\rm S}$ is spin independent since the interaction only occurs between  particles of different spin. Hence, the integrand in Eq.~\eqref{eq:E_Th} behaves as ${\frac{1}{k^2}}$ at high momentum and the action of the regularizing operator gives:
\begin{equation}
\langle H  \rangle = \int \frac{d^D{k}}{(2\pi)^D} \sum_\sigma 
\left[ \frac{\hbar^2 k^2}{2m} n_{{\mathbf k},\sigma} -\frac{\hbar^2 {\rm S} \tilde{a}_D^2}{2m(1+k^2 \tilde{a}_D^2)} \right] , 
\end{equation}
where the use of the length $\tilde{a}_D$ defined in Eq.~\eqref{eq:aDtilde} permits one to gather in a single form the energy theorems for the three different dimensions.

\section{Few-body bound states}

\label{sec:fewbody}

\subsection{Efimov states}

\begin{figure}
\includegraphics[width=8cm,clip]{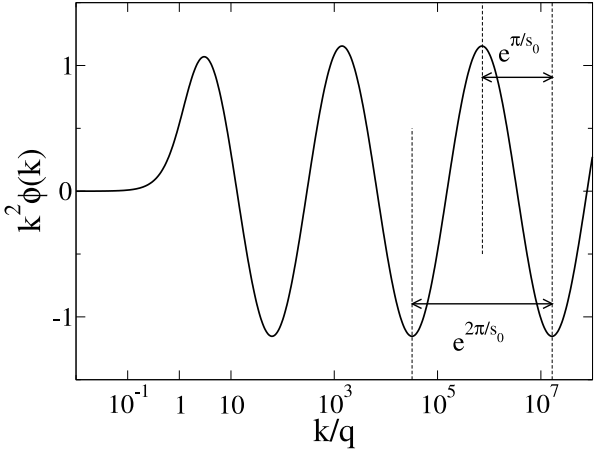}
\caption{Shape of the expession of ${k^2\phi(k)}$ in Eq.~\eqref{eq:eigensol_unitary} plotted in a semi-log plot as a function of $k/q$.}
\label{fig:eigensol_unitary}
\end{figure}

In this section, the generic case where the Efimov effect occurs is considered: three identical bosons of mass $m$ interact in 3D with a pairwise interaction of vanishing range with respect to the 3D scattering length $a_3$ \cite{Efi70,Efi71}. These states have been observed for the first time in ultra-cold atoms \cite{Kra06,Kno09,Zac09,Noa09}.  In what follows, some universal properties of Refs.~\cite{Efi70,Efi71,Bra06} are recovered from the STM equation by using the nodal condition introduced in Ref.~\cite{Pri10b}. However, it is worth pointing out that deviations from universal predictions which are observed in experiments can be taken into account through finite range models \cite{Din09,Lee07,Mas08,Jon10}. The energy of a trimer ${E=E_3}$ in its center-of-mass frame is negative (and less than the dimer's energy $E_2$ if it exists) so that the complementary solution  in Eq.~\eqref{eq:integral_condition} is zero ${(|\Psi^{(0)}\rangle=0)}$. The binding wavenumber $q>0$ of a trimer is defined from:
\begin{equation}
 E_3=-\frac{\hbar^2q^2}{m}.
\end{equation}
The Bose statistics imposes that the source amplitudes for each pair of particle coincide with the same function: 
\begin{equation}
\langle \{{\boldsymbol \xi}^{12}\}|{\rm S}_\Psi^{1 \leftrightharpoons 2} \rangle
=\langle \{{\boldsymbol \xi}^{13}\}|{\rm S}_\Psi^{1 \leftrightharpoons 3} \rangle
=\langle \{{\boldsymbol \xi}^{23}\}|{\rm S}_\Psi^{2 \leftrightharpoons 3} \rangle.
\end{equation}
As a consequence of translation invariance, the source amplitudes in the center-of-mass frame can be written as:
\begin{equation}
\langle \{{\boldsymbol \xi}^{12}\} | {\rm S}_\Psi^{1 \leftrightharpoons 2} \rangle =  
(2\pi)^3 \delta(\mathbf k_1+\mathbf k_2+\mathbf k_3 ) \, {\rm F}({\mathbf k}_3) .
\label{eq:ansatz3B}
\end{equation}
%For convenience, an imaginary collisional wave number for a given pair ${(ij)}$ is defined from its collisional energy ${E_{\rm col}^{(ij)}}$ as:
%\begin{equation}
%E_{\rm col}^{(ij)}({\mathbf k}) = - \frac{\hbar^2q_{\rm col}^2}{m}  \ , \mbox{with}\ q_{\rm col}=\sqrt{ q^2 + \frac{3 k^2}{4}}.
%\end{equation}
The integral eigenequation Eq.~\eqref{eq:integral_condition} takes the form:
\begin{equation}
\frac{{\rm F}({\bf k})}{f_3\left(i\sqrt{ q^2 + \frac{3 k^2}{4}}\right)}= 8\pi \int \frac{d^3{u}}{(2\pi)^3}\, \frac{ {\rm F}({\bf u})}{u^2 +k^2 + {\bf k}.{\bf u} +q^2}.
\label{eq:Skorniakov}
\end{equation}
Equation \eqref{eq:Skorniakov} is the so-called STM equation~\cite{Sko57}. This equation is rotationally invariant and can thus be studied in each momentum sector. In Ref.~\cite{Dan61}, Danilov showed that as it stands Eq.~\eqref{eq:Skorniakov} is ill defined: it supports a continuum of negative energy solutions in the $s$ wave sector of ${F(\mathbf k)}$. Hence, the Bethe-Peierls asymptotic method which is at first sight adapted for modeling the three-boson resonant problem does not permit one to derive a well-defined eigenequation. The $s$ wave component of the source amplitude ${\rm F}({\bf k})$ is denoted as,
\begin{equation}
\int \frac{d\Omega}{4\pi} {\rm F}({\bf k}) = \phi(k),
\end{equation} 
and $\phi(k)$ verifies the integral equation,
\begin{equation}
\frac{\phi(k)} {f_3\left(i\sqrt{ q^2 + \frac{3 k^2}{4}}\right)} = \frac{2}{\pi} \int_0^\infty\!\! du \, \phi(u) {\mathcal K}_q(k,u),
\label{eq:Skorniakov_onde_s}
\end{equation}
where the $s$ wave kernel of the STM equation is given by:
\begin{equation}
{\mathcal K}_q(k,u)=\frac{u}{k} \ln \left(\frac{u^2+k^2+q^2+ku}{u^2+k^2+q^2-ku}\right) .
\end{equation}
The fact that there exists a continuum of negative energy solutions means that the Bethe-Peierls model is not self-adjoint for the three-boson problem. Danilov found a way to restore the self-adjointness by introducing a supplementary condition on the high momentum asymptotic behavior of the function $\phi(k)$ \cite{Dan61}. The method of Danilov is based on the fact that for all values of $q$ and ${a \ne 0}$, Eq.~\eqref{eq:Skorniakov_onde_s} supports solutions with the asymptotic behavior:
\begin{equation}
 \phi(k) \operatornamewithlimits{\sim}_{k \to \infty} A k^{is_0-2} + B k^{-is_0-2} ,
\label{eq:conjugate}
\end{equation}
where $A$ and $B$ are two constants and  ${s_0}$ solves the equation ${\sin(\pi s_0/6)=\sqrt{3}s_0\cos(\pi s_0/2)/8}$  $(s_0\simeq 1.00624\dots)$. The zero-range approach is made self-adjoint if one fixes the asymptotic phase shift between the two conjugate behavior $k^{\pm i s_0}$ for all values of $a$ and $q$. However, Minlos and Fadeev showed that even with this supplementary phase-shift condition the spectrum is not bounded from below \cite{Min62}: this is the so-called Thomas collapse which is characteristic of zero-range forces \cite{Tho35}. In 1970, Efimov solved the three-boson problem in the resonant regime by introducing the notion of a three-body parameter which is also in the zero-range limit, a way to fix the asymptotic phase-shift in Eq.~\eqref{eq:conjugate}. In the $\mathbf k$ representation it is referred in what follows as $\kappa^\star$ and is defined through the asymptotic behavior of ${\phi(k)}$:
\begin{equation}
\phi(k) \operatornamewithlimits{\propto}_{k \to \infty} \frac{1}{k^2} \sin \left[ s_0 \ln \left(\frac{k\sqrt{3}}{\kappa^\star} \right)\right] .
\label{eq:Efi}
\end{equation}
By construction the three-body parameter is not unique [Eq.~\eqref{eq:Efi} is invariant in a change ${\kappa^\star \to \kappa^\star\exp(\pi/s_0)}$]. At unitarity (\emph{i.e.}, ${|a|\to \infty}$ and ${b_3 \to 0}$), Efimov showed that the spectrum of trimers is characterized by an accumulation point at zero energy: binding wavenumbers of the trimers are related the each to the others by a scaling factor,
\begin{equation}
q_n= q_p e^{-(n-p)\pi/s_0} ,
\label{eq:binding_Efi}
\end{equation}
where ${(n,p) \in {\mathbb Z}^2}$. Interestingly, in Ref.~\cite{Gog08} the analytical expression for the source amplitude of the trimers has been found at unitarity:
\begin{equation}
\phi(k)=\frac{1}{k\sqrt{ q^2 + \frac{3 k^2}{4}}} \sin \left[ s_0 \operatornamewithlimits{arcsinh} \left( \frac{k\sqrt{3}}{2q} \right) \right] .
\label{eq:eigensol_unitary}
\end{equation}
In particular, this result shows that the choice made for the definition of ${\kappa^\star}$ in Eq.~\eqref{eq:Efi} is such that the spectrum of the zero-range theory at unitarity is ${q_n=\kappa^\star \exp(n\pi/s_0)}$ with  ${n \in {\mathbb Z}}$. The shape of the function ${k^2 \phi(k)}$ of Eq.~\eqref{eq:eigensol_unitary} is given in Fig.~(\ref{fig:eigensol_unitary}).
For increasing values of ${k}$ starting from ${k=0}$, the first zero of ${\phi(k)}$ is located at the momentum 
${k=\frac{2q}{\sqrt{3}}\operatornamewithlimits{sinh}(\pi/s_0)}$. Using, this property, it is possible to filter a set of solutions of Eq.~\eqref{eq:Skorniakov_onde_s} satisfying Eqs.~(\ref{eq:Efi}) and (\ref{eq:binding_Efi}) for sufficiently high quantum numbers (shallow states) and with a spectrum bounded from below. This is done by imposing a nodal condition on the eigenstates,
\begin{equation}
\phi(k^{\rm reg}_p)= 0 ,
\label{eq:nodal_condition} 
\end{equation}
where the node ${k^{\rm reg}_p}$ is chosen among the set:
\begin{equation}
k^{\rm reg}_p= \frac{\kappa^\star}{\sqrt{3}}e^{p\pi/s_0} \quad p\in {\mathbb Z} .
\label{eq:kreg}
\end{equation}
The position of the node in Eq.~\eqref{eq:kreg} fixes the minimum energy of the spectrum. For example if one chooses the nodal condition for ${p=2}$, using the fact that ${\exp(\pi/s_0) \simeq 2\sinh(\pi/s_0)}$, the minimum energy $E_0$ is almost equal to ${-\hbar^2\kappa^\star\,^2 \exp(2\pi/s_0)/m}$ [with a relative error 
${2\exp(-2\pi/s_0)\simeq 3.8\times10^{-3}}$) and the spectrum at unitarity is asymptotically (\emph{i.e.}, for large values of $n$] given by
\begin{equation}
E_n = -\frac{\hbar^2\kappa^\star\,^2}{m} \exp\left[\frac{-2(n-1)\pi}{s_0}\right] \quad,  n \in {\mathbb N} .
\label{eq:spectrum_cut}
\end{equation}
For a finite scattering length $a_3$ and for ${k|a_3|\gg 1}$, the eigenfunctions ${\phi(k)}$ of Eq.~\eqref{eq:Skorniakov_onde_s} have the same behavior as the unitary solutions of Eq.~\eqref{eq:eigensol_unitary} so that Eq.~\eqref{eq:nodal_condition} can be also used as a filtering condition and permits one to recover the universal spectrum of the zero-range theory for energies much larger than $E_0$. In principle, the zero-range theory also called 'universal theory' in the literature is recovered by imposing the filtering condition at an arbitrary large node $k^{\rm reg}_p$.  To summarize this discussion, the nodal condition in Eq.~\eqref{eq:nodal_condition} has two roles: first, it imposes the Danilov's asymptotic phase shift for states of sufficiently high quantum number and second, it imposes a minimum energy to the spectrum. For a realistic finite range force, the possible values of the binding wavenumbers obtained from the zero-range theory in Eq.~\eqref{eq:binding_Efi} are such that ${q_n b_3\ll 1}$, the ground state energy in Eq.~\eqref{eq:spectrum_cut} has thus to be chosen higher than ${-\hbar^2/(mb_3^2)}$.

The nodal condition in Eq.~\eqref{eq:nodal_condition} applied in Eq.~\eqref{eq:Skorniakov_onde_s} can be transformed into an integral condition:
\begin{equation}
0 = - \frac{2}{ \pi} \int_0^\infty\!\! du \, \phi(u) {\mathcal K}_q(k^{\rm reg}_p,u).
\label{eq:constraint}
\end{equation}
Subtracting Eq.~\eqref{eq:constraint} from Eq.~\eqref{eq:Skorniakov_onde_s} gives a regularized STM equation \cite{Pri10b}:
\begin{multline}
\frac{\phi(k)}{ {f_3\left(i\sqrt{ q^2 + \frac{3 k^2}{4}}\right)}} = \\ \frac{2}{\pi} \int_0^{\infty} \!\! du \, 
\left[{\mathcal K}_q(k,u)-{\mathcal K}_q(k^{\rm reg}_p,u) \right] \phi(u),
\label{eq:sko_regular}
\end{multline}
which encapsulates the nodal condition. The Danilov-Efimov contact condition \eqref{eq:Efi} is implemented \emph{exactly} in the STM equation in the limit where the integer $p$ tends to infinity in Eq.~\eqref{eq:sko_regular}. Numerical solutions of Eq.~\eqref{eq:sko_regular} are obtained by introducing an ultraviolet cut-off ${Q}$ in the integral. One can verify that results are insensitive to the choice made on the cut-off for ${Q \gg k^{\rm reg}_p}$. For instance considering the unitary limit, taking the nodal condition at ${k^{\rm reg}_1}$ and an uv cut-off at ${5\times 10^2 \times \kappa^\star}$, one finds  the relative error for the ground state equals to ${3.8\times 10^{-3} \simeq 2\exp(-2\pi/s_0)}$ and for the first excited states ${(n\le 8)}$ one finds a relative error less than ${10^{-4}}$. In practice, the value of the scaling factor ${\exp(2\pi/s_0) \simeq 515}$ is relatively large so that for the nodal condition at ${k^{\rm reg}_2}$, the spectrum  is very close to the universal spectrum beginning from the second branch ${(n \ge 1)}$ and for an inverse scattering length  ${1/|a_3|}$ smaller or of the order of ${\kappa^\star}$. 

\begin{figure}[hx]
\includegraphics[width=8cm,clip]{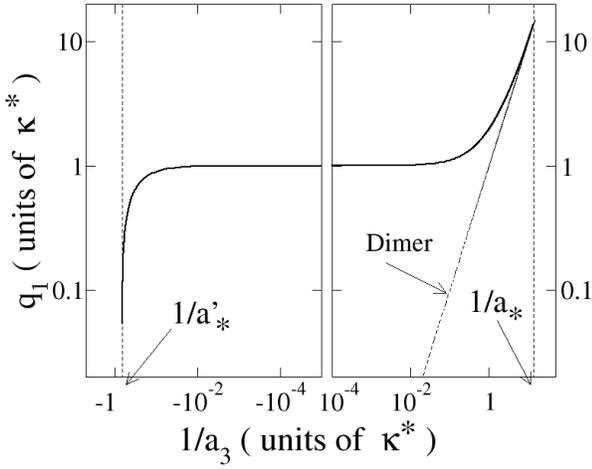}
\caption{Solid line: Second branch of the trimer's spectrum obtained with the nodal condition at ${k_2^{\rm reg}}$. Vertical dashed line: Trimer's appearance thresholds. Oblique dashed line: Atom-dimer continuum limit. The value of the scattering length at the trimer's appearance threshold ${a^*\simeq-1.51/\kappa^\star}$ and ${a'\,^*\simeq .0713/\kappa^\star}$ are close to the results of the ``universal theory'' in Refs.~\cite{Bra06,Gog08}, ${a^*\simeq-1.507/\kappa^\star}$ and ${a'\,^*\simeq.0707/\kappa^\star}$. The difference is due to the fact that the zero-range theory is recovered only for a nodal condition at an arbitrarily large ${k_p^{\rm reg}}$.}
\label{fig:branch}
\end{figure}

\begin{figure}[hx]
\includegraphics[width=8cm,clip]{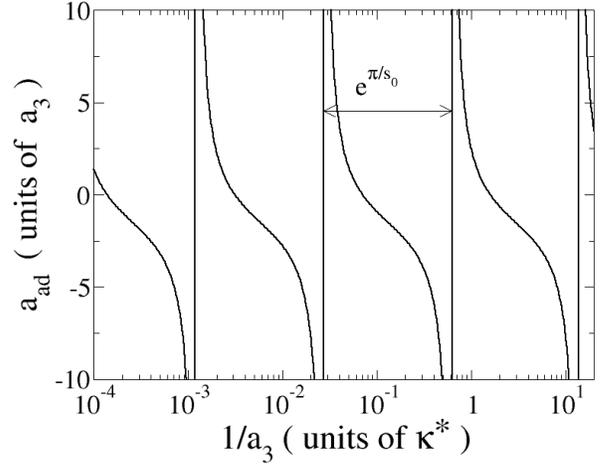}
\caption{Atom-dimer scattering length ${a_{\rm ad}}$ computed with the nodal condition at ${k_2^{\rm reg}}$ and plotted in a semi-log scale as a function of $\kappa^\star/a_3$ for ${a_3 k_2^{\rm reg} \ll 1}$.}
\label{fig:a_ad}
\end{figure}

In Fig.~(\ref{fig:branch}), the second Efimov branch of the regularized STM Eq.~\eqref{eq:sko_regular} with the nodal condition at ${k^{\rm reg}_2}$  has been plotted as a function of ${(1/a_3)}$. As in Ref.~\cite{Bra06}, the thresholds for the appearance of the trimers are denoted by ${a^*}$ at the atom-dimer continuum limit and by ${a'\,^*}$ at the three-atom continuum limit. As a consequence of the choice of the nodal condition at ${k^{\rm reg}_p}$ for a finite $p$, results slightly differ from the zero-range theory in Ref.~\cite{Bra06}.
The atom-dimer scattering problem (regime where ${a_3>0}$) can be also solved by using the same filtering technique. The atom-dimer scattering length is computed as a function of the atomic scattering length ${a_3}$ for a given value of the three-body parameter ${\kappa^\star}$. To this end, the atom-dimer collisional energy is set to zero (${q=1/a_3}$) and the ansatz for the source amplitude is \cite{Sko57}
\begin{equation}
\phi(k)=2\pi^2 \frac{\delta(\mathbf k)}{k^2} + 4\pi \frac{g(k)}{k^2} .
\end{equation}
Using the regularized STM equation one obtains
\begin{multline}
\frac{3 g(k)a_3}{8\left[1+\sqrt{1+\frac{3}{4}(ka_3)^2}\right]} =\frac{a_3^2}{1+(ka_3)^2}-\frac{a_3^2}{1+(k_p^{\rm reg}a_3)^2}\\ 
+ \int_0^{\infty} \!\! \frac{du}{\pi u^2} \, 
\left[{\mathcal K}_{1/a_3}(k,u)-{\mathcal K}_{1/a_3}(k^{\rm reg}_p,u) \right] g(u).
\label{eq:atdim}
\end{multline}
The atom-dimer scattering length denoted  (${a_{\rm ad}}$) is the zero momentum limit of the function  ${g(k)}$ and is plotted in Fig.~(\ref{fig:a_ad}) for the nodal condition taken at ${k_2^{\rm reg}}$.
Figure~(\ref{fig:a_ad}) is limited to atomic scattering lengths ${a_3}$ much smaller than ${1/k_2^{\rm reg}}$ \emph{i.e.}, in a regime where the nodal condition permits to recover the universal theory with high accuracy. The scattering length ${a_{\rm ad}}$ diverges at the threshold of appearance of an Efimov's trimer [${a_3 \simeq a^*\exp(n\pi/s_0)}$, ${n\in \mathbb N}$] and thus exhibits the log-periodicity which is a characteristics of the Efimov physics.

In Ref.~\cite{Bed99} another regularizing technique of the STM equation was derived in the framework of the effective field theory. In this last formulation, the role of the integral counter-term explicitly depends on the uv integral cut-off of the integral ($Q$) in such a way that the three-body parameter has a fixed value for all $Q$. In Refs.~\cite{Ham01,Afn04} a subtraction technique has been also introduced in order to regularize the atom-dimer scattering problem. In these last references, the subtraction is made at zero momentum and therefore imposes the exact value of the atom-dimer scattering length. Thus, this regularization scheme does not correspond to the nodal condition of Eq.~\eqref{eq:nodal_condition}. It is interesting to write down the equation for the atom-dimer scattering amplitude ${g(k)}$ obtained within this scheme:
\begin{multline}
\frac{3 g(k)a_3}{8\left[1+\sqrt{1+\frac{3}{4}(ka_3)^2}\right]} - \frac{3 a_{\rm ad} a_3}{16} =\frac{a_3^2}{1+(ka_3)^2}-a_3^2\\
+ \int_0^{\infty} \!\! \frac{du}{\pi u^2} \, 
\left[{\mathcal K}_{1/a_3}(k,u)-\frac{2u^2a_3^2}{1+u^2 a_3^2} \right] g(u),
\label{eq:atdimEFT}
\end{multline}
and to compare it with Eq.~\eqref{eq:atdim}.

\subsection{Heteronuclear trimers}

As a consequence of interesting predictions \cite{Pet03,Kar07a,Kar07b,Pet07,Pri10a,Nis08,Lev09,Nis09,Ors10,Hel10,Cas10}, heteronuclear systems play an important role in ultra-cold physics. In this section, the eigenequation for heteronuclear trimers is considered without external potential. However, it is worth noticing that an external potential can deeply change the physical properties of the system as it is the case for example in systems of mixed dimensions \cite{Nis09,Lev09}. Particles (labeled by $i$) can be either bosons or fermions and have a mass ${m_i}$ which can take two possible values: ${m_i=m}$ or ${m_i=M}$. Bosons (respectively fermions) of mass $M$ are denoted by $B$ (respectively, by $F$) while bosons (respectively, fermions) of mass $m$ are denoted by $b$ (respectively, by  $f$). Fermionic particles are supposed to have two possible internal states denoted $\uparrow$ and $\downarrow$. The configurations studied here are thus constructed from the set of particles ${\{b,B,f_\uparrow,f_\downarrow,F_\uparrow,F_\downarrow\}}$. The present study is also restricted to '2+1' few-body systems where the two-body interaction is non-vanishing for heterogeneous pairs of particles only. As an example in the case of ${(bF_\uparrow)}$ interacting pairs, other pairs like $(bF_\downarrow)$ or ${(bB)}$ are not interacting. For fermions neglecting ${(F_\uparrow F_\uparrow)}$ or ${(f_\uparrow f_\uparrow)}$ interactions is an exact assumption which follows from the Pauli principle. However, neglecting ${(BB)}$ or ${(bb)}$ interactions is not an exact hypothesis and only means that these interactions are negligible with respect to a heteronuclear interaction. Thus the true inter-atomic forces of the heteronuclear interacting pairs are in the vicinity of a $s$ wave resonance while interaction for pairs of identical bosons is neglected. For convenience, the following notations are introduced for the different combinations of particles masses ${m_i}$:
\begin{eqnarray}
&&M_{(ij)}= m_i + m_j \\
&&\mu_{i,(jk)}= \frac{m_i M_{(jk)}}{m_i+M_{(jk)}}\\
&&\mu_{(ij),(kl)}=\frac{M_{(ij)} M_{(kl)}}{M_{(ij)}+M_{(kl)}} .
\end{eqnarray}
In this section, two identical atoms of mass $M$ (particles 1 and 2) interact with another one of mass $m$ (particle 3). All the possible bound states can be thus deduced from the two configurations: ${(BBb)}$ [or equivalently ${(BBf)}$] and ${(F_\uparrow F_\uparrow b)}$ [or equivalently ${(F_\uparrow F_\uparrow f)}$]. 
\begin{figure}[h]
\includegraphics[width=8cm,clip]{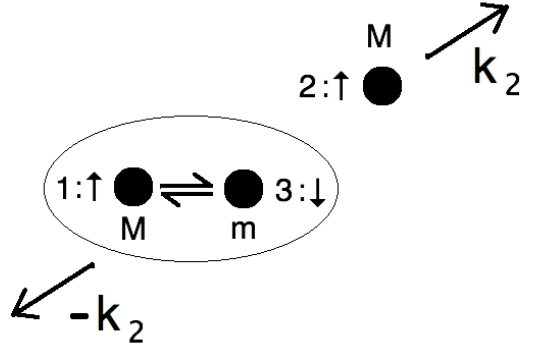}
\caption{Schematic representation of the momentum coordinates used for the source amplitude associated with the contact condition between particles 1 and 3.}
\label{fig:bille3}
\end{figure}
An eigenequation for trimers is obtained for a negative energy $E=E_3$, 
\begin{equation}
E_3 = -\frac{\hbar^2 q^2}{2\mu_{2,3}} < 0 ,
\end{equation}
and if a dimer of binding energy $E_2$ exists then ${E_3<E_2}$. In the center-of-mass frame, the source amplitude associated with the pair ${\rm (23)}$ is:
\begin{equation}
\langle \{{\boldsymbol \xi}^{23}\} | {\rm S}_\Psi^{2 \leftrightharpoons 3} \rangle =  
(2\pi)^D \delta(\mathbf k_1+\mathbf k_2+\mathbf k_3 ) \, {\rm F}({\mathbf k}_1)  .
\label{eq:ansatz3body}
\end{equation}
The other source amplitude with the momentum coordinates represented schematically in Fig.~(\ref{fig:bille3}), is deduced from Eq.~\eqref{eq:ansatz3body} by using the exchange symmetry between particle $1$ and  particle $2$:
\begin{equation}
\langle \{{\boldsymbol \xi}^{13}\} | {\rm S}_\Psi^{1 \leftrightharpoons 3} \rangle = s_{13}
(2\pi)^D \delta(\mathbf k_1+\mathbf k_2+\mathbf k_3 ) \, {\rm F}({\mathbf k}_2),
\end{equation}
where $s_{13} =1$ when particles $1$ and $2$ are bosons and $s_{13}=-1$ if they are fermions. For simplifying the notations, it is useful to introduce the mass ratio $y$ defined by
\begin{equation}
y=\frac{\mu_{2,3}}{m_3}=\frac{M}{M+m} .
\end{equation}
Assuming that the  $\epsilon \to 0$ limit is well defined, the eigenequation is obtained from the contact condition for the pair  ${(2 : M ; 3 :m )}$ in Eq.~\eqref{eq:integral_condition} where ${|\Psi^{(0)}\rangle = 0}$:
\begin{equation}
\frac{-s_{13} \hbar^{2} {\rm F}({\bf k})}{ 2\mu_{2,3} T_D(E^{\rm col}_{\mathbf k})} \\
= \int \frac{d^D{u}}{(2\pi)^D}\, \frac{ {\rm F}({\bf u})}{u^2 +k^2 + 2 y {\bf k}.{\bf u} +q^2}.
\label{eq:3b}
\end{equation}
In Eq.~\eqref{eq:3b}  the identity ${d^D k_{23}=d^D k_2}$ has been used (the contact condition is performed at a fixed value of ${\mathbf K_{23}}$) and ${E^{\rm col}_{\mathbf k}}$ is the collisional energy:
\begin{equation}
E^{\rm col}_{\mathbf k} = E_3 - \frac{\hbar^{2} k^2}{2\mu_{1,(23)}} <0 .
\end{equation}
A detailed study of the three-body bound states in 2D has been performed in Ref.~\cite{Pri10a}, and in this subsection the discussion is centered on the 3D case. By using the rotational symmetry of the kernel in Eq.~\eqref{eq:3b}, one can fix an arbitrary direction: $\hat{e}_{\mathbf q}$ and expand the source amplitude ${\rm F}({\bf k})$ in terms of partial waves as:
\begin{equation}
{\rm F}({\bf k}) = \sum_{l=0}^\infty P_l(\hat{e}_{\mathbf k}\cdot\hat{e}_{\mathbf q}) F_l(k) .
\end{equation}
\begin{figure}[t]
\includegraphics[width=8cm,clip]{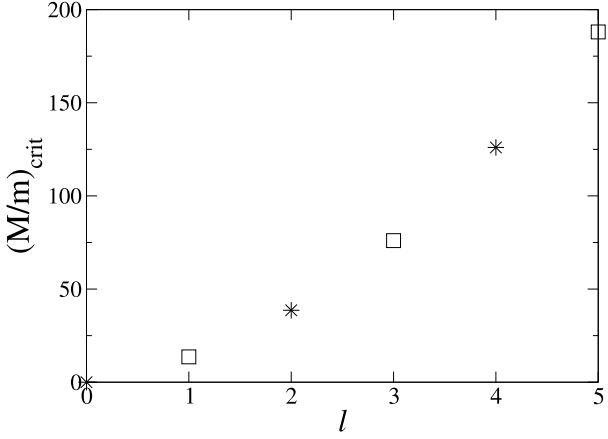}
\caption{Critical mass ratio ${\left(\frac{M}{m}\right)^{\rm crit}_l}$ for the threshold of appearance of an Efimov spectrum in each partial wave $l$ of Eq.~\eqref{eq:3bMm}.}
\label{fig:threshold}
\end{figure}
Each component $F_l(k)$ verifies the integral equation
\begin{equation}
\frac{\hbar^{2}s_{13} (-1)^{l+1} F_l(k)}{ 2\mu_{1,3}T_3(E^{\rm col}_{\mathbf k})} =  
\int_0^\infty\!\! du \, \frac{u F_l(u)}{y \pi k} Q_l\left(\frac{{u^2+k^2+q^2}}{2yku}\right),
\label{eq:3bMm}
\end{equation}
where $Q_l$ is a Legendre function of the second kind. One can notice that the standard STM Eq.~\eqref{eq:Skorniakov} is obtained by setting formally $s_{13}=2$ and $y=1/2$ in Eq.~\eqref{eq:3bMm}. For a sufficiently large value of the mass ratio, the Efimov effect appears in non zero partial waves \cite{Efi73,Bul75}. As in the three-boson case, in this regime  and in the limit of large momentum $k$ Eq.~\eqref{eq:3bMm} supports a pair of solutions ${F_l(k) \sim k^{-2\pm is}}$ (${s\in \mathbb{R}}$). The critical values of the mass ratio of the Efimov theshold in the $l$ wave [denoted ${\left(\frac{M}{m}\right)^{\rm crit}_l}$] are plotted in Fig.~(\ref{fig:threshold}). They have been deduced from Eq.~\eqref{eq:3bMm} in the Appendix \ref{appendixA} and the results found with this method coincide with the ones computed in Refs.~\cite{Bul75,Pet04a,Kar07b}.

\subsection{Heteronuclear tetramers}

\begin{figure}
\includegraphics[width=8cm,clip]{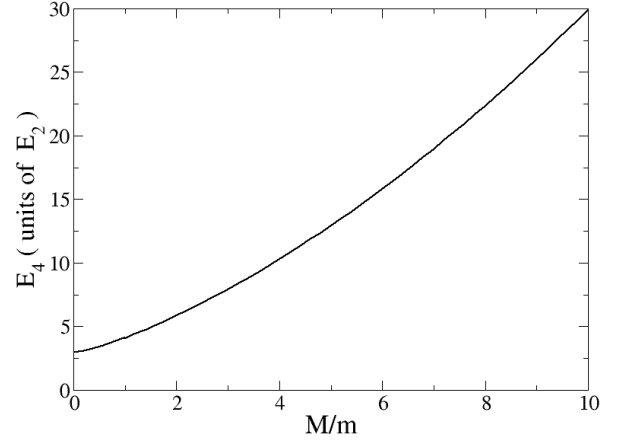}
\caption{Ground state branch for the three bosons of mass $M$ interacting with one impurity of mass $m$ in 2D as a function of the mass ratio.}
\label{fig:3Bp1}
\end{figure}

Recent theoretical \cite{Plat04,Yam06,Bro06,Ham07,Ste09,Cas10} and experimental \cite{Fer09,Pol09} progress has been achieved in the four-body problem with ultra-cold atoms. In this section, a zero-range eigenequation for four-particle bound states  or ``tetramers'' is considered. The binding energy is denoted $E_4$ and
\begin{equation}
E_4=-\frac{\hbar^2 q^2}{2\mu_{1,2}},
\end{equation}
where ${E_4<E_3}$ and/or ${E_4<2 E_2}$ if a trimer and/or dimer exists. A particle $(i)$ is characterized by a momentum ${{\mathbf k}_i}$ and a mass ${m_i}$ equals to $m$ or $M$. In the center-of-mass frame, the source amplitude associated with the pair ${(12)}$ can be written as:
\begin{equation}
\langle \{{\boldsymbol \xi}^{12}\} | {\rm S}_\Psi^{1 \leftrightharpoons 2} \rangle = (2\pi)^D \delta(\sum_{n=1}^4 {\mathbf k}_n ) 
 {\rm F}({\boldsymbol K}_{12},{\boldsymbol k}_{34}) .
\label{eq:ansatz4body}
\end{equation}
\begin{figure}[h]
\includegraphics[width=8cm,clip]{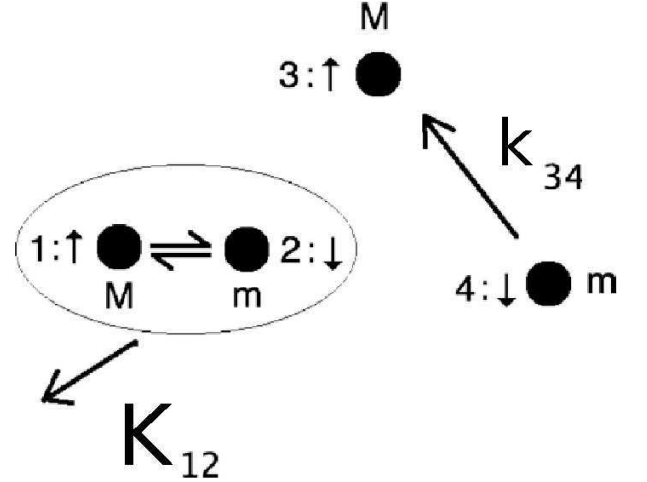}
\caption{Schematic representation of the momentum coordinates used for the source amplitude associated with the contact condition between particles 1 and 2 for a system of fermions with two-mass components.}
\label{fig:bille2p2}
\end{figure}
The integral equation satisfied by the function $F$ is obtained from Eq.~\eqref{eq:integral_condition} (here ${(ij)=(1,2)}$). A specific configuration is represented schematically in Fig.~(\ref{fig:bille2p2}). The summation on the left hand side of the integral equation is composed of source terms which are deduced from Eq.~\eqref{eq:ansatz4body} by using the permutation symmetry and the statistics of the particles. In order to have a general equation for the different possible configurations, a statistical factor denoted ${s_{np}}$ is introduced for each pair of particles. Depending on the system, for two interacting particles $n$ and $p$, the statistical factor is ${s_{np}=\pm 1}$ or ${s_{np}=0}$ for noninteracting particles. Using this notation, the different source amplitude are written in Tab.~(\ref{tab:sources}).
\begin{table}
\begin{tabular}{|c|c|c|}
\hline
Pair $({ij)}$  & Permutations & ${\displaystyle \langle \{{\boldsymbol \xi}^{ij}\} | {\rm S}_\Psi^{i \leftrightharpoons j} \rangle}$\\
\hline 
${(12)}$ &  &  ${\displaystyle  (2\pi)^D \delta(\sum_{n=1}^4 {\mathbf k}_n ) \, {\rm F}({\boldsymbol K}_{12},{\boldsymbol k}_{34})}$\\
\hline 
${(13)}$ & ${2 \leftrightarrow 3 }$ &   ${\displaystyle  s_{13} (2\pi)^D \delta(\sum_{n=1}^4 {\mathbf k}_n ) \, {\rm F}({\boldsymbol K}_{13},{\boldsymbol k}_{24})}$\\
\hline
${(14)}$ &  ${2 \leftrightarrow 4 }$ &   ${\displaystyle  s_{14} (2\pi)^D \delta(\sum_{n=1}^4 {\mathbf k}_n ) \, {\rm F}({\boldsymbol K}_{14},{\boldsymbol k}_{32})}$\\
\hline
${(23)}$ &  ${1 \leftrightarrow 3 }$ &   ${\displaystyle  s_{23} (2\pi)^D \delta(\sum_{n=1}^4 {\mathbf k}_n ) \, {\rm F}({\boldsymbol K}_{23},{\boldsymbol k}_{14})}$\\
\hline
${(24)}$ &  ${1 \leftrightarrow 4 }$ &   ${\displaystyle  s_{24} (2\pi)^D \delta(\sum_{n=1}^4 {\mathbf k}_n ) \, {\rm F}({\boldsymbol K}_{24},{\boldsymbol k}_{31})}$\\
\hline
${(34)}$ &  ${1 \leftrightarrow 3 }$,  ${2 \leftrightarrow 4}$ &   ${\displaystyle  s_{34} (2\pi)^D \delta(\sum_{n=1}^4 {\mathbf k}_n ) \, {\rm F}({\boldsymbol K}_{34},{\boldsymbol k}_{12})}$\\
\hline
\end{tabular}
\caption{Source terms for each pair $(ij)$ deduced by using the permutation symmetry of the four-body wave function from the source term of the interacting pair $(12)$ in Eq.~\eqref{eq:ansatz4body}. The statistical factors $s_{ij}$ are given in Table~\ref{tab:configuration}.}
\label{tab:sources}
\end{table}
Different possible configurations of the two-mass component system and the corresponding statistical factors are listed in Table \ref{tab:configuration}.
\begin{table}
%\begin{ruledtabular}
\begin{tabular}{|c|c|c|}
\hline
Configuration & ${(m_1,m_2,m_3,m_4)}$ &   ${(s_{13},s_{14},s_{23},s_{24},s_{34})}$   \\
\hline
${BBBB}$ &  ${(m,m,m,m)}$ &    ${(1,1,1,1,1)}$   \\\hline 
${(f\, {\rm or}\,  b)BBB}$  &   ${(m,M,M,M)}$ &  ${(1,1,0,0,0)}$ \\ \hline   
${BbBb}$ &  ${(M,m,M,m)}$ &    ${(0,1,1,0,1)}$   \\\hline 
${F_\uparrow f_\downarrow F_\uparrow f_\downarrow}$ &  ${(M,m,M,m)}$ & ${(0,-1,-1,0,1)}$ \\\hline
${B f_\uparrow B f_\uparrow}$  &  ${(M,m,M,m)}$ & ${(0,-1,1,0,-1)}$ \\ \hline
${(f\, {\rm or}\,  b) F_\uparrow F_\uparrow F_\uparrow }$  &   ${(m,M,M,M)}$ &  ${(-1,-1,0,0,0)}$ \\ \hline   
\end{tabular}
%\end{ruledtabular}
\caption{Different possible 4-body configurations. The notation is as follows: $B$ (respectively $F$) means that the atom is a boson (respectively fermion) of mass $M$, $b$ (respectively $f$) means that the atom is a boson (respectively a fermion) of mass $m$. The fermions have two possible internal states $\uparrow$ and $\downarrow$. For each configuration, interaction is non vanishing only between one type of heterogeneous pair. The statistical factors $s_{ij}$ appears in the integral equation~\eqref{eq:integral_4b}.}
\label{tab:configuration}
\end{table}
Equation~\eqref{eq:integral_condition} can be simplified to
\begin{multline}
 \int \frac{d^D{{k}_{12}}}{(2\pi)^D} 
\operatornamewithlimits{\sum_{n<p}}_{\rm p>2}
s_{np} 
\frac{ {\rm F}\left({\boldsymbol K}_{np},{\boldsymbol k}_{kl} \right)}{\kappa^2 + k_{12}^2 }
=\frac{-\hbar^{2}  {\rm F}({\mathbf u},{\mathbf v})} {2\mu_{1,2}T_D(E_{\rm col}^{(12)})} ,
\label{eq:contact_4b}
\end{multline}
where ${\mathbf u}={\boldsymbol K}_{12}$, ${\mathbf v} = {\boldsymbol k}_{34}$. In Eq.~\eqref{eq:contact_4b} the momentum $\kappa$ is related to the collisional energy by
${E_{\rm col}^{(12)}=\hbar^2 \kappa^2/(2\mu_{1,2})}$, thus:
\begin{equation}
\kappa^2= q^2 + \frac{\mu_{1,2}}{\mu_{(12),(34)}} u^2 +  \frac{\mu_{1,2}}{\mu_{3,4}} v^2.
\end{equation}
Using the notations defined in Table \ref{tab:new_coordinates}, the integral equation which encapsulates the Bethe-Peierls contact condition can finally be written as: 
\begin{multline}
\int \frac{d^D{k}}{(2\pi)^D} 
\left[ 
 s_{13} \frac{{\rm F }({\mathbf k},{\mathbf v}_{13})}{\kappa^2+\sigma_{13}^2}
+ s_{14} \frac{{\rm F }({\mathbf k},{\mathbf v}_{14})}{\kappa^2+\sigma_{14}^2} 
+ s_{23} \frac{{\rm F }({\mathbf k},{\mathbf v}_{23})}{\kappa^2+\sigma_{23}^2}
\right.
\\
\left.
+ s_{24} \frac{{\rm F }({\mathbf k},{\mathbf v}_{24})}{\kappa^2+\sigma_{24}^2}+s_{34} \frac{ {\rm F }( -{\mathbf u},{\mathbf k})}{\kappa^2 + k^2} \right]
= \frac{-\hbar^{2} {\rm F}({\mathbf u},{\mathbf v})} { 2\mu_{1,2}T_D(E_{\rm col}^{(12)})},
\label{eq:integral_4b}
\end{multline}
\begin{table}
%\begin{ruledtabular}
\begin{tabular}{|c|c|c|}
\hline
Pair $({ij)}$  &  ${{\mathbf v}_{ij}}$ \  & ${{\boldsymbol \sigma}_{ij}}$   \\
\hline 
${(13)}$ &  
${\displaystyle \frac{m_4 {\mathbf u}}{M_{3,4}} + {\mathbf v} - \frac{m_4 {\mathbf k}}{M_{2,4}}}$   &  
${\displaystyle \left(\frac{\mu_{1,2}}{m_1} - \frac{\mu_{3,4}}{m_3}\right) \, {\mathbf u} -{\mathbf v} +{\mathbf k}}$ 
\\
\hline 
${(14)}$ & 
${\displaystyle -\frac{m_3 {\mathbf u}}{M_{3,4}} + {\mathbf v} + \frac{m_3 {\mathbf k}}{M_{2,3}}}$  &
${\displaystyle \left(\frac{\mu_{1,2}}{m_1} - \frac{\mu_{3,4}}{m_4}\right) \, {\mathbf u} + {\mathbf v} +{\mathbf k}}$ \\
\hline
${(23)}$ &  ${\displaystyle \frac{m_4 {\mathbf u}}{M_{3,4}} + {\mathbf v} + \frac{m_4 {\mathbf k}}{M_{1,4}}}$  & 
${\displaystyle  \left(\frac{\mu_{1,2}}{m_1} - \frac{\mu_{3,4}}{m_4}\right) \, {\mathbf u} + {\mathbf v} - {\mathbf k}}$
\\
\hline
${(24)}$ &  ${\displaystyle \frac{m_3 {\mathbf u}}{M_{3,4}} - {\mathbf v} + \frac{m_3 {\mathbf k}}{M_{1,3}}}$ &
${\displaystyle \left(\frac{\mu_{1,2}}{m_1} - \frac{\mu_{3,4}}{m_3}\right) \, {\mathbf u} - {\mathbf v} - {\mathbf k}} $
\\  
\hline
\end{tabular}
%\end{ruledtabular}
\caption{Coordinates appearing in Eq.~\eqref{eq:integral_4b}.}
\label{tab:new_coordinates}
\end{table}
In the particular case where the system is composed of two heterogeneous dimers made of fermions (${F_\uparrow f_\downarrow F_\uparrow f_\downarrow}$), one recovers the same integral equation as in Eq.~(16) of Ref.~\cite{Pet05}. In 3D depending on the statistics of the particle and of the mass ratio, Eq.~\eqref{eq:integral_4b} is not in general well defined as a consequence of the Efimov-Thomas effect. For systems composed of four particles in 2D, bound states of particles of same mass has been considered in \cite{Bro06}. In the present work, the binding energies of 2D ground tetramers are computed numerically as a function of the mass ratio of the interacting particles in the $BBBb$ configuration. Calculations are restricted to $s$ wave tetramers by using the ansatz
\begin{equation}
{\rm F}({\mathbf u},{\mathbf v})=F(u,v,\theta)  \quad \mbox{with} \ \theta=\angle{({\mathbf u},{\mathbf v})} .
\end{equation}
In order to check the numerical computation,  the particular case of four identical bosons which has been already obtained by several authors \cite{Plat04,Bro06} has been considered. In this configuration, two bound states have been found with binding energies which are close to already published results: ${E_4/E_2=197}$ and ${24}$, to be compared with ${E_4/E_2=197.3}$ and ${25.5}$ in Ref.~\cite{Plat04} or with ${E_4/E_2=194}$ and $24$ in Ref.~\cite{Bro06}. Results for three identical bosons interacting resonantly with another particle is shown in Fig.~(\ref{fig:3Bp1}).
In 3D for the same $bBBB$ configuration, similarly to the three-boson original STM equation,  Eq.~\eqref{eq:integral_4b} does not constitute a well-defined problem. The nature and properties of eigenstates of this system in the zero-range limit are yet unsolved. The ${b F_\uparrow F_\uparrow F_\uparrow}$ or ${f_\downarrow F_\uparrow F_\uparrow F_\uparrow}$ configurations have been studied recently in Ref.~\cite{Cas10} and a pure four-body Efimov effect (\emph{i.e.}, without a three-body Efimov state) involving a four-body parameter has been found for a mass ratio of ${13.384<M/m<13.607}$. 

\section{CONCLUSIONS}

In this paper, it has been shown how the zero-range approach can be formulated for different problems in a unified framework  without technical intricacy for all dimensions ${D\le 3}$. In the few-body problem, general integral eigenequations are obtained straightforwardly in the momentum representation. Efimov effect is deduced from the Skorniakov Ter-Martirosian equation by using a nodal condition and a subtracting technique which is close to regularizing schemes used in the Effective Field Theory. 

Various open issues remain to be solved for few-body systems in the limit of zero-range forces. For example, the question whether or not the 3-body parameter is enough for describing 4-bosons properties in 3D is still under debate \cite{Ste09,Ham07}. For heteronuclear few-body systems, taking into account the external trapping (which is a natural concern in regards to experiments) makes the problem even richer. The external potential differs for different atomic species and this degree of freedom allows one to explore realistic intermediate situations between the fully three-dimensional homogeneous configuration and the limit of mixed dimensions where interesting predictions have been already made in the three-body case \cite{Nis09,Lev09}. The present formalism can be also used to model systems where interaction occurs between three spin components like in $^6$Li experiments \cite{Ott08,Wil09,Huc09,Lom10}. Another open issue concerns the possible occurrence of Efimov states for more than four particles in 3D. The four-body Efimov phenomenon has been found to occur for a specific interval of mass ratio in the 3+1 fermionic problem in Ref.~\cite{Cas10}. Therefore, one can wonder whether or not a pure Efimov effect in a $N+1$ fermionic system can also occur and at which critical mass ratio for $N\ge 3$. Despite the difficulty of this problem, the integral equation of the system in the limit of zero-range forces can be easily deduced from the present formalism and is given below as a concluding remark. Each of the $N$ polarized fermions of mass $M$ labeled by $i$ (${1\le i\le N}$) interact with only one impurity (particle ${i=N+1}$) of mass $m$. In the center-of-mass frame, the source amplitude for the pair ${N,(N+1)}$ can be written as:
\begin{multline}
\langle \{{\boldsymbol \xi}^{N,(N+1)}\}|{\rm S}_\Psi^{N \leftrightharpoons N+1}\rangle=(2\pi)^D\delta\left(\sum_{i=1}^{N+1}\mathbf k_i \right)\\
\times F(\mathbf k_1,\mathbf k_2\dots \mathbf k_{N-1}),
\end{multline}
and other source amplitudes ${|{\rm S}_\Psi^{i \leftrightharpoons N+1}\rangle}$ (${i=1\cdots N-1}$)  are deduced from this ansatz by using the fermionic statistics. The general integral equation for this problem which encapsulates the Bethe-Peierls asymptotic condition is then obtained from Eq.~\eqref{eq:integral_condition}: 
\begin{widetext}
\begin{multline}
\int \frac{d^D k_N}{(2\pi)^D} 
\frac{F(\mathbf k_N,\mathbf k_2,\mathbf k_3, \cdots \mathbf k_{N-1})+F(\mathbf k_1,\mathbf k_N, \mathbf k_3, \cdots \mathbf k_{N-1})+\cdots + F(\mathbf k_1,\mathbf k_2, \cdots \mathbf k_{N-2}, \mathbf k_{N})}
{-E -i 0^+ +\frac{\hbar^2}{2M} \sum_{i=1}^N k_i^2 + \frac{\hbar^2}{2m} \left( \sum_{i=1}^{N} \mathbf k_i \right)^2 }\\
=\frac{F(\mathbf k_1, \mathbf k_2, \cdots \mathbf k_{N-1})}{T_D(E_{\rm col}+i0^+)}-\int \frac{d^D{k}_{N}}{(2\pi)^D} \Psi^{(0)}(\mathbf k_1,\cdots \mathbf k_N),
\end{multline}
\end{widetext}
where ${E_{\rm col}=E-\frac{\hbar^2}{2M} \sum_{i=1}^{N-1} k_i^2 - \frac{\hbar^2}{2(M+m)} \left( \sum_{i=1}^{N-1} \mathbf k_i \right)^2}$, $\Psi^{(0)}$  is a complementary solution associated with the free Hamiltonian at energy $E$ in the center-of-mass frame (it is equal to zero for ${E<0}$) and $F$ is antisymmetric under the exchange of two coordinates ${(\mathbf k_i,\mathbf k_j)}$. 

\section*{ACKNOWLEDGMENTS}

It is a pleasure to thank Y. Castin, M. Jona-Lasinio, C. Mora and F. Werner for discussions. LPTMC is UMR 7600 of the CNRS and its Cold Atoms group is associated with IFRAF.

\appendix

\section{THRESHOLDS FOR THE '1+2'-BODY EFIMOV EFFECT}

\label{appendixA}

In this appendix, the values of the mass ratio for which an Efimov effect occurs are deduced from Eq.~\eqref{eq:3bMm}. To this end, zero energy solutions at unitarity ${|a_3|=\infty}$ are considered (solutions for finite energy have the same high momentum behavior). In this regime, Eq.~\eqref{eq:3bMm} is scale invariant and this allows one to search for power law solutions: ${F_l(k)=k^{\nu+l-2}}$, where $\nu$ is a function of the mass ratio. For convenience, the following dimensionless parameter $t$ is introduced:
\begin{equation}
t = \arcsin(y) = \arcsin\left(\frac{M}{M+m}\right). 
\end{equation}
Thus, the scattering amplitude is ${f_3(\kappa_{\mathbf k})=1/(k \cos t)}$ and Eq.~\eqref{eq:3bMm} gives for each partial wave $l$ an eigenvalue equation: ${\lambda(\nu,l,t)=0}$ where,
\begin{equation}
\lambda(\nu,l,t) = \cos t - \frac{s_{13} (-1)^l}{\pi \sin t } \int_0^\infty dz z^{\nu+l-1} Q_l\left(\frac{1+z^2}{2 z \sin t}\right) .
\label{eq:lambdanul}
\end{equation}
The integrand in Eq.~\eqref{eq:lambdanul} is positive, hence for a Bose (respectively Fermi) statistics  only even (respectively odd) values of $l$ can support a solution. For ${x>1}$, ${Q_l(x)=\frac{1}{2} P_l(x) \ln(\frac{x+1}{x-1})+W_l(x)}$ where $W_l(x)$ is a polynomial of order $l$. The polynomial $W_l$ does not contribute in the integral of Eq.~\eqref{eq:lambdanul} and ${\lambda(\nu,l,t)}$ can be thus expressed in terms of a sum of functions of the form:
\begin{equation}
I(\gamma,t) = \int_0^\infty dz\, z^{\gamma-1} \ln \left(\frac{z^2 + 2 z \sin t +1}{z^2 - 2z \sin t +1} \right),
\label{eq:Igamma}
\end{equation}
with ${0 \le t \le \frac{\pi}{2}}$ and ${|\Re(\gamma)|<1}$. In Eq.~\eqref{eq:Igamma}, ${I(\gamma,t)}$ can be calculated explicitly as: 
\begin{equation}
I(\gamma,t) = \frac{2\pi  \sin (\gamma t)}{\gamma \cos (\frac{\pi \gamma}{2})}.
\end{equation}
In the $l$ wave, Efimov states are characterized by a power law such that ${\nu=-l+is}$, where $s$ is real. For example, for two identical bosons interacting with another particle (${s_{13}=1}$), in the $s$ wave sector,
\begin{equation}
\lambda\left(\nu,l=0, t \right)= \cos t - \frac{\sin (\nu t )}{\nu \sin(t) \cos \left(\frac{\pi\nu}{2}\right)} .
\label{eq:Efi_swave}
\end{equation}
Equation~\eqref{eq:Efi_swave} admits solution of the form ${\nu=is}$ for all values of the mass ratio and ${s\to 0}$ for ${t\to 0}$. For higher $l$ waves, Efimov states appear above a critical value of the mass ratio. The threshold is obtained by searching the value of the parameter ${t=t^{\rm crit}_l}$ such that:
\begin{equation}
\lim_{\nu \to -l} \lambda \left( \nu,l,t^{\rm crit}_l \right) = 0 .
\end{equation}
In the $p$-wave sector which concerns the case of two identical fermions interacting with another particle (${s_{13}=-1}$) one gets
\begin{equation}
\lambda\left(\nu,l=1, t \right)= \cos t - \frac{(\nu+2) \sin (\nu t ) - \nu \sin [(\nu+2) t]}{2\nu(\nu+2)\sin^2 t \cos t 
\cos \left(\frac{\pi\nu}{2}\right)} ,
\end{equation}
and the threshold is obtained from the equation
$
{t^{\rm crit}_1 = \tan ( t^{\rm crit}_1) - \frac{\pi}{2} \sin^2( t^{\rm crit}_1)},
$
which gives the critical mass ratio ${\left(\frac{M}{m}\right)^{\rm crit}_{l=1} \simeq 13.6}$ found in Refs.~\cite{Bul75,Pet03}. For higher partial waves, the critical values of the mass ratio ${\left(\frac{M}{m}\right)^{\rm crit}_l}$ found by this method coincide with the ones computed in Ref.~\cite{Kar07b} where the hyperspherical method was used. Results are reported in Fig.~(\ref{fig:threshold}).

In the regime where an Efimov effect occurs, the regularizing technique can be achieved by fixing the asymptotic phase shift which is a function of the three-body parameter (denoted $\kappa^\star_l$) in the partial wave $l$:
\begin{equation}
F_l(k) \operatornamewithlimits{\propto}_{k \to \infty} \frac{1}{k^2} \sin \left[ s \ln \left(\frac{k\sqrt{3}}{\kappa^\star_l} \right)\right] .
\label{eq:Efi2}
\end{equation}
A set of solutions satisfying Eq.~\eqref{eq:Efi2} can be filtered from Eq.~\eqref{eq:3bMm} by imposing the nodal condition:
\begin{equation}
F_l(k^{\rm reg}_{p,l})=0 \quad \mbox{where} \quad k^{\rm reg}_{p,l} = \frac{\kappa^\star_l}{\sqrt{3}} e^{p \pi/s}\, , p \in {\mathbb N}.
\label{eq:filtre2} 
\end{equation}
Exact zero-range theory is obtained in the limit where ${p \to \infty}$. Universal results are recovered if the node ${k^{\rm reg}_{p,l}}$ is chosen at a large value as compared to all the low energy scales (${k^{\rm reg}_{p,l} \gg 1/|a|,q}$). Using the subtracting scheme of the three-boson problem, Eq.~\eqref{eq:filtre2} can be also incorporated into the integral equation Eq.~\eqref{eq:3bMm}:
\begin{multline}
\frac{\hbar^{2}  F_l(k)}{ 2\mu_{1,3}T_3(E^{\rm col}_{\mathbf k})} = - \frac{s_{13} (-1)^l}{y \pi} 
\int_0^\infty\!\! du \, u F_l(u) \\
\left[ 
\frac{1}{k} Q_l\left(\frac{{u^2+k^2+q^2}}{2yku}\right)-
\frac{1}{k_{p,l}^{\rm reg}} Q_l\left(\frac{u^2+(k_{p,l}^{\rm reg})^2+q^2}{2yuk_{p,l}^{\rm reg}}\right)
\right]
.
\label{eq:3bregular}
\end{multline}


\begin{thebibliography}{0}
\expandafter\ifx\csname natexlab\endcsname\relax\def\natexlab#1{#1}\fi
\expandafter\ifx\csname bibnamefont\endcsname\relax
  \def\bibnamefont#1{#1}\fi
\expandafter\ifx\csname bibfnamefont\endcsname\relax
  \def\bibfnamefont#1{#1}\fi
\expandafter\ifx\csname citenamefont\endcsname\relax
  \def\citenamefont#1{#1}\fi
\expandafter\ifx\csname url\endcsname\relax
  \def\url#1{\texttt{#1}}\fi
\expandafter\ifx\csname urlprefix\endcsname\relax\def\urlprefix{URL }\fi
\providecommand{\bibinfo}[2]{#2}
\providecommand{\eprint}[2][]{\url{#2}}

\end{thebibliography}


\begin{thebibliography}{99}

%Interaction n-p for the 'Diplon' ie H2
\bibitem{Bet35} H. Bethe and R. Peierls, Proc. R. Soc. London, Ser. {\bf A} {\bf 148}, 146 (1935).

%'The Scattering of Slow Neutrons by Bounds Protons'
%Lien entre le pseudo-potentiel de Fermi et la condition de contact
\bibitem{Bre47} G. Breit, Phys. Rev. {\bf 71}, 215 (1947).

\bibitem{Bla52} J.M. Blatt and V.F. Weisskopf, in {\it 'Theoretical Nuclear Physics'}, Wiley, New York (1952). 
%Eqs 3.36-3.37: first appearance of the final expression of the 
%zero-range pseudopotential for $s$ wave scattering
%effective range approximation used in particular for neutron-proton cross section: 

%Atomic Scattering in the Presence of an External Confinement and a Gas of Impenetrable Bosons
\bibitem{Ols98} M. Olshanii, Phys. Rev. Lett. {\bf 81}, 938 (1998).

%Bose-Einstein Condensation in Quasi-2D Trapped Gases
\bibitem{Pet00} D. S. Petrov, M. Holzmann, and G. V. Shlyapnikov, Phys. Rev. Lett. {\bf 84}, 2551 (2000).

%Interatomic collisions in a tightly confined Bose gas
\bibitem{Pet01} D. S. Petrov and G. V. Shlyapnikov, Phys. Rev. A {\bf 64}, 012706 (2001).

%Analytical solutions for the dynamics of two trapped interacting ultracold atoms
\bibitem{Idz06} Z. Idziaszek, T. Calarco, Phys. Rev. A {\bf 74}, 022712 (2006).

%Three-body problem in Fermi gases with short-range interparticle interaction
\bibitem{Pet03} D. S. Petrov, Phys. Rev. A {\bf 67}, 010703R (2003).

%Weakly Bound Dimers of Fermionic Atoms
\bibitem{Pet04a} D.S. Petrov, C. Salomon, and G. V. Shlyapnikov, Phys. Rev. Lett. {\bf 93}, 090404 (2004).

%Exact scaling transform for a unitary quantum gas in a time dependent harmonic potential
\bibitem{Cas04} Y. Castin, C. R. Phys. {\bf 5}, 407 (2004).

%One particle in a box: The simplest model for a Fermi gas in the unitary limit
\bibitem{Pri04a} L. Pricoupenko, Y. Castin, Phys. Rev. A {\bf 69}, 051601(R) (2004).

%Diatomic molecules in ultracold Fermi gases-novel composite bosons
\bibitem{Pet05} D.S. Petrov, C. Salomon and G.V. Shlyapnikov, J. Phys. B Mol. Opt. Phys. {\bf 38}, S645 (2005).

%Confinement-induced resonances for a two-component ultracold atom gas in arbitrary quasi-one-dimensional traps
\bibitem{Pea05} V. Peano, M. Thorwart, C. Mora, R. Egger, New J. Phys. {\bf 7}, 192 (2005).

%Unitary gas in an isotropic harmonic trap: Symmetry properties and applications
\bibitem{Wer06a} F. Werner and Y. Castin, Phys. Rev. A {\bf 74}, 053604 (2006).

%Unitary Quantum Three-Body Problem in a Harmonic Trap
\bibitem{Wer06b} F. Werner and Y. Castin, Phys. Rev. Lett. {\bf 97}, 150401 (2006).

%Universal description of the rotational-vibrational spectrum of three particles with zero-range interactions
\bibitem{Kar07a} O. I. Kartavtsev, A. V. Malykh, Pis'ma ZhETF {\bf 86}, 713 (2007).

%Low-energy three-body dynamics in binary quantum gases
\bibitem{Kar07b} O. I. Kartavtsev, A. V. Malykh, J. Phys. B: At. Mol. Opt. Phys. {\bf 40}, 1429 (2007).

\bibitem{pwave} This assertion is not true in the peculiar case of fully polarized identical fermions where $s$ wave scattering is forbidden as a consequence of the Pauli principle. In this paper, scattering in non zero partial wave is considered to be non resonant and is thus neglected.

\bibitem{Fes62} H. Feshbach, Annals of Physics {\bf 19}, 287 (1962).

%Feshbach Resonances in Ultracold Gases
\bibitem{Chi10} C. Chin, R. Grimm, P. Julienne, and E. Tiesinga, Rev. Mod. Phys. {\bf 82}, 1225 (2010).

\bibitem{QGLD} Proceedings of the School ``Quantum Gases in Low Dimensions'', J. Phys. IV 
{\bf 116} (EDP Sciences, Les Ulis-France, 2004).

%Observation of a One-Dimensional Tonks-Girardeau Gas
\bibitem{Kin04a} T. Kinoshita, T. Wenger, and D.S. Weiss, 
%\href{http://www.sciencemag.org/cgi/content/abstract/sci;305/5687/1125}
{Science {\bf 305}, 1125 (2004)}.
%DOI: 10.1126/science.1100700

%Tonks Girardeau gas of ultracold atoms in an optical lattice
\bibitem{Par04} B. Paredes,  A. Widera, V. Murg, O. Mandel, S. F\"{o}lling, I.~Cirac, G.V.~Shlyapnikov, T.W.~H\"{a}nsch, and I. Bloch, 
%\href{http://www.nature.com/nature/journal/v429/n6989/abs/nature02530.html}
{Nature (London) {\bf 429}, 277 (2004)}.
%doi:10.1038/nature02530

%Many-body physics with ultracold gases
\bibitem{Blo08} I. Bloch, J. Dalibard, W. Zwerger, 
%\href{http://link.aps.org/doi/10.1103/RevModPhys.80.885}
{Rev. Mod. Phys. {\bf 80}, 885 (2008)}.
%ISSN: 0034-6861
%URL:http://link.aps.org/doi/10.1103/RevModPhys.80.885
%DOI: 10.1103/RevModPhys.80.885 

%Two-dimensional Bose fluids: An atomic physics perspective
%in\textgravedbl {\sl Lecture notes from the Varenna Summer School 23 June-3 July 2009},
\bibitem{Had09} Z. Hadzibabic, J. Dalibard, Proceedings of the International School of Physics 'Enrico Fermi', Course CLXXIII:
{\sl 'Nano optics and atomics: transport of light and matter waves'}, Varenna 2009, edited by R. Kaiser and D. Wiersma (IOS
Press, Amsterdam, 2011).

%Lambda potential
\bibitem{Ols02} M. Olshanii and L. Pricoupenko, Phys. Rev. Lett.{\bf 88}, 010402 (2001).

%Short-Distance Correlation Properties of the Lieb-Liniger System and Momentum Distributions of Trapped One-Dimensional Atomic Gases
\bibitem{Ols03} M. Olshanii, V. Dunjko, Phys. Rev. Lett. {\bf 91}, 090401 (2003).

%Energetics of a strongly correlated Fermi gas
\bibitem{Tan08a} S. Tan, Ann. Phys. N.Y., {\bf 323}, 2971 (2008).

%Large momentum part of a strongly correlated Fermi gas
\bibitem{Tan08b} S. Tan, Ann. Phys. N.Y., {\bf 323}, 2987 (2008).

%Particle distribution tail and related energy formula
\bibitem{Com09}  R. Combescot, F. Alzetto, and X. Leyronas, Phys. Rev. A {\bf 79}, 053640 (2009).
%URL:http://link.aps.org/doi/10.1103/PhysRevA.79.053640
%DOI:10.1103/PhysRevA.79.053640

%Exact relations for quantum-mechanical few-body and many-body problems with short-range interactions in two and three dimensions
\bibitem{Wer10} F. Werner, Y. Castin,  {\it arXiv:1001.0774}.

\bibitem{Sko57} G. Skorniakov and K. Ter-Martirosian, Sov. Phys. JETP {\bf 4}, 648 (1957).

\bibitem{Tho35} L. H. Thomas, Phys. Rev. {\bf 47}, 903 (1935).

\bibitem{Pri10b} L. Pricoupenko, Phys. Rev. A {\bf 82}, 043633 (2010). 

%Universality in few-body systems with large scattering length
\bibitem{Bra06} E. Braaten, H.-W. Hammer, Phys. Rep. {\bf 428}  259 (2006).

%Nonsingular integral equation for two-body scattering and applications in two and three dimensions
\bibitem{Sto88} H.T.C. Stoof, L.P.H. de Goey, W.M.H.M. Rovers, P.S.M. Kop Jansen, and B.J. Verhaar, Phys. Rev. A {\bf 38}, 1248 (1988).

%Effective interactions, Fermi–Bose duality, and ground states of ultracold atomic vapors in tight de Broglie waveguides
\bibitem{Gir04} M. Girardeau, H. Nguyen, M. Olshanii, Opt. Comm. {\bf 243}, 3 (2004).

\bibitem{Convention_f2D} In Ref.~\cite{Pri08b} a different convention was taken for ${f_2(k_0)}$.

%Resonant scattering of ultracold atoms in low dimensions
\bibitem{Pri08b} L. Pricoupenko, Phys. Rev. Lett. {\bf 100}, 170404 (2008).

%Variational approach for the two-dimensional trapped Bose Einstein condensate
\bibitem{Pri04b} L. Pricoupenko, Phys. Rev. A {\bf 70}, 013601 (2004).

%Pseudopotential in resonant regimes
\bibitem{Pri06} L. Pricoupenko, Phys. Rev. A {\bf 73}, 012701 (2006).

%Paper on the neutron-proton interaction
\bibitem{Fer36} E. Fermi, Ric. Sci. {\bf 7-II}, 13 (1936).
%See Eqs.(71)-(103) and especially Eq.~(80): first appearance of delta_R

%Resonances in ultracold collisions of 6Li, 7Li, and 23Na
\bibitem{Moe95} A.J. Moerdijk, B.J. Verhaar and A. Axelsson, Phys. Rev. A {\bf 51}, 4852 (1995).

\bibitem{Pet04b} D.S. Petrov, Phys. Rev. Lett. {\bf 93}, 143201 (2004). 

%Atom-Atom Scattering under Cylindrical Harmonic Confinement: Numerical and Analytic Studies of the Confinement Induced Resonance
\bibitem{Ber03} T. Bergeman, M.G. Moore and M. Olshanii, Phys. Rev. Lett. {\bf 91}, 163201 (2003).

\bibitem{Gel64} I. Gelfand, G. Shilov, {\it Generalized Functions} (Academic Press, New York, 1964), vol. {\bf 1},.

\bibitem{Sch78} L. Schwartz, {\it Th\'eorie des distributions} (Hermann, Paris, 1978).

\bibitem{Gra94} I. Gradshteyn and I. Ryzhik, in {\it Table of Integrals, Series, and Products}, 5th ed. (Academic Press, San Diego, 1994), Sec. {\bf 9.511}.

\bibitem{Meromorphic} A function ${f(s)}$ which is defined in the domain ${\Re(s)>1}$ by  ${f(s)=\int_0^\infty dt\, t^{s-2} g(t)}$ where ${g(0)}$ and ${g'(0)}$ are finite, can be written after integration by parts as ${f(s)=\int_0^\infty dt\, t^{s-1} g'(t) /(1-s)}$, where the function ${g}$ is supposed to decrease sufficiently rapidly to zero at ${t=\infty}$ for ensuring the convergence of the integrals.  This last relation is well defined for ${0<\Re(s)}$ and ${s \ne 1}$, it is the meromorphic continuation of the first expression in this domain of the variable $s$ and it coincides also with its finite part.

\bibitem{Efi70} V. Efimov, Phys. Lett. B {\bf 33}, 563 (1970).

\bibitem{Efi71} V. Efimov, Sov. J. Nucl. Phys. {\bf 12}, 589 (1971).

%Evidence for Efimov quantum states in an ultracold gas of caesium atoms; Accepted 2 February 2006
\bibitem{Kra06} T. Kraemer, M. Mark, P. Waldburger, J. Danzl, C. Chin, B. Engeser, A. Lange, K. Pilch, A. Jaakkola, H.-C. N\"{a}gerl and R. Grimm, Nature  (London) {\bf 440}, 315 (2006).

%Observation of an Efimov-like resonance in ultracold atom-dimer scattering
\bibitem{Kno09} S. Knoop, F. Ferlaino, M. Mark, M. Berninger, H. Schoebel, H.-C. Naegerl, R. Grimm, Nat. Physics {\bf 5}, 227 (2009). 

%Observation of an Efimov spectrum in an atomic system
\bibitem{Zac09} M. Zaccanti, B. Deissler, C. D'Errico, M. Fattori, M. Jona-Lasinio, S. M\"{u}ller, G. Roati, M. Inguscio and G. Modugno, Nat. Phys. {\bf 5}, 586 (2009).

%Observation of Universality in Ultracold 7Li Three-Body Recombination
\bibitem{Noa09} N. Gross, Z. Shotan, S. Kokkelmans, and L. Khaykovich, Phys. Rev. Lett. {\bf 103}, 163202 (2009).

%The short-range three-body phase and other issues impacting the observation of Efimov physics in ultracold quantum gases
\bibitem{Din09} J. D'Incao, C. H. Greene, B. D. Esry, J. Phys. B {\bf 42}, 044016 (2009). 

%Excited Thomas-Efimov levels in ultracold gases
\bibitem{Lee07} M.D. Lee, T. K\"{o}hler and P.S. Julienne, Phys. Rev. A {\bf 76}, 012720 (2007). 

%Efimov states near a Feshbach resonance
\bibitem{Mas08} P. Massignan and H.T.C. Stoof, Phys. Rev. A {\bf 78}, 030701(R) (2008).

%Three Resonant Ultra-Cold Bosons: Off-Resonance Effects"
\bibitem{Jon10} M. Jona-Lasinio and L. Pricoupenko, Phys. Rev. Lett. {\bf 104}, 023201 (2010). 

\bibitem{Dan61} G. Danilov, Sov. Phys. JETP {\bf 13}, 349 (1961).

\bibitem{Min62} R. Minlos and L. Faddeev, Sov. Phys. JETP {\bf 14}, 1315 (1962).

%Analytical solution of the bosonic three-body problem
\bibitem{Gog08} A.O. Gogolin, C. Mora, R. Egger, Phys. Rev. Lett. {\bf 100}, 140404 (2008). 

%Renormalization of the Three-Body System with Short-Range Interactions
\bibitem{Bed99} P.F. Bedaque, H.-W. Hammer, and U. van Kolck, Phys. Rev. Lett. {\bf 82}, 463 (1999).

%Three-body problem with short-range forces: Renormalized equations and regulator-independent results
\bibitem{Ham01} H.-W. Hammer and T. Mehen, Nucl. Phys. A {\bf 690}, 535 (2001).
%e-Print: nucl-th/0011024

\bibitem{Afn04} I.R. Afnan and D.R. Phillips, Phys. Rev. C {\bf 69}, 034010 (2004).
%e-Print: nucl-th/0312021.

%Crystalline Phase of Strongly Interacting Fermi Mixtures
\bibitem{Pet07} D.S. Petrov, G.E. Astrakharchik, D.J. Papoular, C. Salomon, and G.V. Shlyapnikov, Phys. Rev. Lett. {\bf 99}, 130407 (2007).

%Universal 1+2-body bound states in planar atomic waveguides 
\bibitem{Pri10a} L. Pricoupenko and P. Pedri, Phys. Rev. A {\bf 82}, 033625 (2010).

%Universal Fermi gases in mixed dimensions
\bibitem{Nis08} Y. Nishida and S. Tan, Phys. Rev. Lett. {\bf 101}, 170401 (2008).

%Atom-Dimer Scattering and Long-Lived Trimers in Fermionic Mixtures
\bibitem{Lev09} J. Levinsen, T.G. Tiecke, J.T.M. Walraven, and D.S. Petrov, Phys. Rev. Lett. {\bf 103}, 153202 (2009).

%Confinement-induced Efimov resonances in Fermi-Fermi mixtures
\bibitem{Nis09} Y. Nishida and S. Tan, Phys. Rev. A {\bf 79}, 060701(R) (2009).

%Three-body problem in heteronuclear mixtures with resonant interspecies interaction
\bibitem{Hel10}  K. Helfrich, H.-W. Hammer, D.S. Petrov, Phys. Rev. A {\bf 81}, 042715 (2010).

%Luttinger liquid of trimers in Fermi gases with unequal masses
\bibitem{Ors10} G. Orso, E. Burovski, T. Jolicoeur, Phys. Rev. Lett. {\bf 104}, 065301 (2010) 
%arXiv:0907.1533

%Four-body Efimov effect
\bibitem{Cas10} Y. Castin, C. Mora, L. Pricoupenko, Phys. Rev. Lett. {\bf 105}, 223201 (2010).

\bibitem{Efi73} V. Efimov, Nucl. Phys. A {\bf 210}, 157 (1973).

\bibitem{Bul75} A. Bulgac and V. Efimov, Sov. Jour. Nucl. Phys. {\bf 22}, 296 (1975).

%Universal Properties of the Four-Boson System in Two Dimensions
%DOI 10.1007/s00601-004-0065-z
\bibitem{Plat04} L. Platter, H.-W. Hammer, and U.-G. Mei{\ss}ner, Few-Body Systems {\bf 35}, 169 (2004).

%Exact diagrammatic approach for dimer-dimer scattering and bound states of three and four resonantly interacting particles
\bibitem{Bro06}  I.V. Brodsky and M.Y. Kagan, A.V. Klaptsov, R. Combescot and X. Leyronas, Phys. Rev A {\bf 73}, 032724 (2006).

%Four-boson scale near a Feshbach resonance
\bibitem{Yam06} M. T. Yamashita, L. Tomio, A. Delfino and T. Frederico, 
%\href{http://iopscience.iop.org/0295-5075/75/4/555/pdf/0295-5075_75_4_555.eps}
{Europhys. Lett. {\bf 75}, 555 (2006)}.
%DOI: 10.1209/epl/i2006-10141-6

%Universal Properties of the Four-Body System with Large Scattering Length
\bibitem{Ham07} H.-W. Hammer, L. Platter, Eur. Phys. J. A {\bf 32}, 113 (2007).

%Signatures of universal four-body phenomena and their relation to the Efimov effect
\bibitem{Ste09} J. von Stecher, J. D'Incao, and C. Greene, Nature Phys. {\bf 5}, 417 (2009).

%%%%

%Evidence for Universal Four-Body States Tied to an Efimov Trimer
\bibitem{Fer09} F. Ferlaino, S. Knoop, M. Berninger, W. Harm, J.P. D'Incao, H.C. N\"{a}gerl, and R. Grimm, Phys. Rev. Lett. {\bf 102}, 140401 (2009).

%Universality in Three- and Four-Body Bound States of Ultracold Atoms
\bibitem{Pol09} S. Pollack , D. Dries, R. Hulet, Science {\bf 326}, 1683 (2009).

%Expérience avec les trois états du Lithium

%Collisional Stability of a Three-Component Degenerate Fermi Gas
\bibitem{Ott08} T. B. Ottenstein, T. Lompe, M. Kohnen, A. N. Wenz, and S. Jochim, Phys. Rev. Lett. 101, 203202 (2008).

%Evidence for an Excited-State Efimov Trimer in a Three-Component Fermi Gas
\bibitem{Wil09} J. R. Williams, E. L. Hazlett, J. H. Huckans, R. W. Stites, Y. Zhang, and K. M. O'Hara, Phys. Rev. Lett. {\bf 103},
130404 (2009).

%Radio-Frequency Association of Efimov Trimers
\bibitem{Lom10} T. Lompe, T. B. Ottenstein, F. Serwane, K. Viering, A. N. Wenz, G. Z\"{u}rn, S. Jochim, Science {\bf 330}, 940 (2010).

%stabilité des systèmes à 3 composantes
%Three-Body Recombination in a Three-State Fermi Gas with Widely Tunable Interactions
\bibitem{Huc09} J. H. Huckans, J. R. Williams, E. L. Hazlett, R. W. Stites, and K. M. O’Hara, Phys. Rev. Lett. {\bf 102}, 165302 (2009).

\end{thebibliography}
\end{document}